\documentclass[12pt]{iopart}

\usepackage{graphicx}
\expandafter\let\csname equation*\endcsname\relax
\expandafter\let\csname endequation*\endcsname\relax
\usepackage{amsmath}
\usepackage{amssymb}
\usepackage{svg}
\usepackage{indentfirst}
\usepackage{xcolor}
\usepackage[hidelinks, hypertexnames=false]{hyperref} 
\usepackage[all]{hypcap}
\usepackage{slashed}
\usepackage{graphicx}
\usepackage{latexsym}
\usepackage{epstopdf}
\usepackage{enumitem}
\usepackage{multirow}
\usepackage{mathtools}
\usepackage{bbm}
\usepackage{bm}
\usepackage{amsfonts}

\usepackage{calligra}
\usepackage{calrsfs}
\usepackage[normalem]{ulem}
\usepackage[USenglish,american]{babel}
\usepackage{braket}
\usepackage{cancel}
\usepackage{physics}
\graphicspath{{./figures/}}

\makeatletter
\begin{document}

\title[Short-wavelength mesophases in ground-states in 2D]{Short-wavelength mesophases in the ground states of core-softened particles in two-dimensions}

\author{Rômulo Cenci, Lucas Nicolao and Alejandro Mendoza-Coto}

\address{Departamento de Física, Universidade Federal de Santa Catarina, 88035-972 Florianópolis, Brazil}
\ead{lucas.nicolao@ufsc.br}
\ead{alejandro.mendoza@ufsc.br}
\vspace{10pt}
\begin{indented}
\item[]December 2025
\end{indented}

\begin{abstract}
We describe the formation of short-wavelength mesophases in a two-dimensional core-softened particle system. By proposing a series of specific ansatz for each relevant phase, we performed a variational analysis to obtain the ground-state phase diagram. Our results reveal a variety of cluster lattice phases with distinct cluster orientations, alongside traditional two-dimensional Bravais lattices such as square, triangular, oblique, and rectangular structures, as well as other non-Bravais arrangements including honeycomb and kagome phases. We characterize in detail the ground-state phase transitions and identify coexistence regions between competing phases, capturing both first-order and continuous transitions. In addition, we highlight the crucial role of the competing length scales introduced by the hard-core repulsion in shaping the rich landscape of mesophases, emphasizing the interplay between intra-cluster structure and inter-cluster organization. {Finally, our analytical results are confronted with extensive molecular dynamics simulations, which interestingly show the existence of decagonal and dodecagonal quasicrystalline phases in regions of the phase diagram that exhibit a high degree of frustration.}
This study provides a systematic framework that could support future investigations of classical thermal melting behavior or quantum phase transitions in similar cluster-forming systems.
\end{abstract}

\section{Introduction}

The spontaneous emergence of spatially modulated phases constitutes a central topic in the study of complex many-body systems, encompassing a wide range of physical contexts. These phases often arise from the presence of competing interactions, which introduce frustration into the system~\cite{Seul1995, Imperio2004}. In classical systems, such frustration can give rise to a variety of modulated phases \cite{Glaser2007, Imperio2006}, including \textit{cluster phases}~\cite{Likos2007, Lenz2012, Diaz2015} termed also ``{clump}''~\cite{Klein1994, Reichhardt_2003} or ``{bubble}'' phases~\cite{mendoza2019}; \textit{inverted-cluster phases}~\cite{Pini2017}, also known as ``{inverse micelle}'' \cite{Glaser2007}, ``{hole}'' \cite{Chacko2015} or ``{anticlump}'' phases \cite{Reichhardt2010}; and \textit{stripe phases}~\cite{Malescio2003, Osterman_2007, Mendoza_2015}. Such phases have been observed experimentally in systems as diverse as block copolymers \cite{Christopher2000}, magnetic thin films \cite{Saratz2010, Xiuzhen2012}, vortex matter~\cite{Cirac2013} and colloidal systems \cite{Menath2021, Ciarella2021, Grillo2020}.
In the last case, even more exotic phases like sigma \cite{Dijkstra2017}, quasicrystaline~\cite{Dotera2014, Zu2017} and cluster-quasicrystaline~\cite{Barkan2014} phases have been observed in simulations of model systems.

While such structures arise in classical systems, in quantum systems the interplay between competing interactions and quantum fluctuations gives rise to even more exotic phases, including supersolids, as is the case of dipolar Bose-Einstein condensates observed both experimentally~\cite{Chomaz2019, Bottcher2019, Norcia2021} and  theoretically~\cite{Ripley2023, Lima2025}, as is the case of soft-core bosons that display quantum cluster crystals phases~\cite{Cinti2014,Cinti_2014_1,Cinti2019,Mendoza_2021_2}, where superfluidity may coexist with non-trivial spatial order \cite{Boninsegni2011, Boninsegni2012b}. In this context, quantum quasicrystaline phases can be observed at finite temperature~\cite{Pupillo2020}, and also at the ground-state level, either by mean of an external potential~\cite{Zampronio_2024} or by careful tuning the pairwise interactions~\cite{Mendoza_2022, gross2025}, giving rise to exotic quantum states of matter \cite{Gross_2024, Mendoza2025}.

Among the class of interactions that drive self-assembly, ultra-soft repulsive potentials, characterized by a bounded core and a negative minimum in their Fourier transform, are responsible for the formation of cluster-crystalline states with particles that might overlap~\cite{Likos2007, Neuhaus_2011, Mendoza_2021}. Typically these potentials produce the stabilization of cluster phases with an integer $n$ occupation number and isostructural transitions as the density is increased~\cite{Mello_2023}. Notably, these are first order phase transitions, that occur through the development of coexistence regions of phases with $n$ and $n+1$ occupancy numbers. Hence, at the beginning of the coexistence region, droplets of the $n+1$ cluster phase nucleates and, as the density increases, these droplets grow until the system presents a pure $n+1$ cluster phase and the coexistence region ends.

When a hard-core repulsion or additional sharp length scale is introduced at the short distances, the tendency to form overlapping clusters can be significantly inhibited, paving the way for the formation of new modulated states {with properties between solid and liquid phases, which are sometimes called {\it mesophases}}~\cite{Glaser2007}. At higher densities, such modulated phases includes long-wavelength stripes, clusters and holes~\cite{Imperio2004,Reichhardt2010,Chacko2015,Caprini2018}. At lower densities, however, these phases are characterized by short-wavelength modulations \cite{Jagla1999, Cacciuto2011, Meng2014, Zhao2017, Somerville2020}, and includes phases such as square lattices \cite{Prestipino2012}, as well as stripes of one particle width, kagome and honeycomb phases \cite{Dijkstra2017, Cardoso2021, Fomin2021, Puccinelli2025}, among many others. In particular, the cluster phases with a small occupation number $n$ are often identified as dimers for $n=2$ \cite{Nogueira2022}, trimers ($n=3$), tetramers ($n=4$) \cite{Meng_2017}, and so forth. {More exotic phases can be found in such systems, as is the case of sigma phases \cite{Zu2017, Dijkstra2017}, considered as periodic approximant for quasicrystalline phases, as well as quasicrystalline phases with $Q$-fold rotational symmetry, with the more common $Q=12$ corresponding to dodecagonal quasicrystal \cite{Zu2017, Dijkstra2017, Dotera2014, Schoberth2016}.}

The mechanism for the pattern formation stems from the competition between the ultra-soft component, which favors clustering, and the hard-core component, which penalizes such configurations~\cite{Caprini2018, Mambretti2021}. 
This combination of soft and hard-core repulsions is sometimes referred as {\it core-softened} or {\it core-corona} potentials as well as {\it hard-core/soft-shoulder} (HCSS) potentials \cite{Jagla1998, Jagla1999, Glaser2007}.
As a result, systems may settle into phases where the stabilized structure arises from the balance of competing interactions rather than simple clustering — a behavior observed in both classical and quantum contexts~\cite{Abreu2022}. 

In situations when the effective hard-core size is much lower than the spacing of the lattice formed by the arrangement of the clusters themselves, the particles belonging to a cluster develop an internal structure that minimizes the hard-core repulsion energy - typically a {BCC or FCC lattice in three dimensions~\cite{Mello_2023, Prestipino_2014} or a triangular lattice in two dimensions}. Meanwhile, in this case, the long-distance physical properties of the cluster lattice remains essentially unaltered in comparison to the corresponding soft-core particle system~\cite{Caprini2018}. This kind of behavior naturally disappears as soon as the size of the clusters becomes comparable to the distance between them. In such regime the intra and inter cluster interactions are not well separated by an energy scale and, consequently, the behavior of the system becomes much richer, which can be explored already at the ground state level \cite{Fornleitner2010}. This scenario is exactly the one explored in Refs. \cite{Mambretti2021} and \cite{Pini2024}, where the authors investigated the properties of a self-assembled crystalline dimer phase in a regime where the hard-core size of the particles is comparable to the distance between dimers forming the lattice. Interestingly, the authors show that the classical ground state is given by a ``nematic'' state in which dimers are oriented in a non-trivial direction with respect to the dimers lattice.  

Within this context, our work aims to present a systematic study of the self-assembling process in a 2D soft-core cluster forming model frustrated by a hard-core interaction. In order to do so, we consider a two-dimensional {model with a core-softened potential that combines the {generalized exponential model} with exponent 4 (GEM-4) and an inverse power-law potential with exponent 6 -- a similar core-softened potential was considered in a previous study \cite{Caprini2018} that investigated cluster-crystal ground-states in much higher densities and much lower effective hard-core sizes than those considered in the present work. We compute in detail the classical ground-state phase diagram, using density and the relative strength of the two repulsive terms as running parameters. Our results show that the hard-core interaction significantly modifies the soft-core scenario of clusters with increasing occupancy. The maximum cluster occupancy decreases rapidly with the hard-core size, while a plethora of new single-particle phases emerge, which are not present in the pure GEM-4 model. Furthermore, we explore the intermediate regime where both interactions act on comparable scales, making the orientational order of the clusters a non-trivial feature that varies along the phase diagram.}

The paper is organized as follows. { In Sect.~\ref{model} we introduce the model and the methods used to determine the classical ground state. Sect.~\ref{results} presents a detailed description of the stabilized phases and their location in the phase diagram, first from direct energy minimization of pure phases and then allowing for phase coexistence; this section also includes a correction to the phase diagram based on simulations.} Finally, Sect.~\ref{conclusions} is devoted to present our final remarks and conclusions.

\section{Model and Methodology}
\label{model}

\subsection{Core-softened repulsive potential model}

{We consider a two-dimensional system of classical particles interacting via a pair potential composed of two competing repulsive terms: an ultrasoft and bounded interaction that favors particle clustering, and a short-range interaction that penalizes particle overlap. It can be generally expressed as:
\begin{equation}
    U(r) = \epsilon_s \exp\left(-(r/R)^\alpha\right) + \epsilon_s (r_0/r)^\gamma,
    \label{pot_dim}
\end{equation}
where $\epsilon_s$ controls the energy scale, and the first term represents the soft repulsion with typical length scale $R$, while the second term corresponds to the power-law hard-core repulsion with typical length scale $r_0$, characterizing the effective particle size. By expressing energies in units of $\epsilon_s$ and distances in units of $R$, we rewrite the potential in the following dimensionless form, which is used throughout this work:
\begin{equation}
    V(r)=\exp\left(-r^\alpha\right)+\frac{C}{r^\gamma},
    \label{pot}
\end{equation}
with the particular values $\alpha=4$ and $\gamma=6$. The dimensionless parameter $C = (r_0/R)^\gamma$ controls the relative strength of the hard-core repulsion. In our analysis and phase diagrams, we often use the parameter $\ell_C \equiv -\log C$, as it conveniently represents the relative hard-core strength across different orders of magnitude. }
The first term in (\ref{pot}) represents the soft part of the pairwise potential, known as the generalized exponential model (GEM-$\alpha$), which is a bounded potential allowing for the formation of cluster-crystal ground states, where clusters are arranged on a triangular lattice and each cluster is formed by overlapping particles, with the occupation number increasing with density for $\alpha>2$~\cite{Prestipino_2014,Neuhaus_2011,Mello_2023}. This ultra-soft repulsion is combined with an inverse-power-law potential, the second term in (\ref{pot}), which effectively acts as a hard-core repulsion, preventing particles from overlapping and thereby enabling the emergence of a variety of fine-structured ground states, partially characterized in this work.

Since the parameter $C$ controls the relative strength of the hard-core repulsion with respect to the ultra-soft interaction, it thereby regulates the different spatial scales on which these two interactions act. For very small $C$ (in the limit $C\to 0$), the hard-core repulsion acts on very short length scales, and the ground state of overlapping particle clusters is replaced by one in which the clusters acquire a small size, composed of particles arranged in a triangular lattice, with the distance between clusters being much larger than the distance between particles within each cluster~\cite{Caprini2018}. In the opposite limit, for $C \gtrsim 0.16$ {($\ell_C \lesssim 0.80$, as shown by our results in the following section in Fig.~\ref{fig:diag})}, the ultra-soft repulsion is completely dominated by the hard-core potential and the ground-state becomes the familiar triangular lattice.

In this work, we are interested in the intermediate regime where both repulsive interactions act on comparable scales and promote complex mesophases. For example, we expect the formation of ground states with a regular lattice of clusters, where the distance between clusters is comparable to the distance between particles inside a cluster, so that the internal structure and orientation of each cluster affect the arrangement of the overall lattice. It is precisely in this regime that mesophases proliferate, giving rise to an intricate phase diagram. 

\subsection{{Molecular dynamics simulations}}

In order to identify candidates for possible short-wavelength ground-states, we performed extensive molecular dynamics simulations. To efficiently explore the configuration space and overcome metastable states, we employed a combination of parallel tempering and simulated annealing techniques. These exploratory simulations were instrumental in identifying the relevant phases and constructing the specific ansatz for our variational analysis. 

{The simulations were performed in a NVT or canonical ensemble, where we control the number of particles, the volume (area) and temperature of the system. The contact with a thermal reservoir in the dynamics of the particles is accomplished through the following Langevin equation in dimensionless form:
\begin{equation}
    \mathbf{\ddot r}_i =
     -\sum_{j\neq i} \mathbf{\nabla}_i V(\mathbf{r}_i-\mathbf{r}_j)
     - \gamma \mathbf{\dot r}_i + \sqrt{2\gamma T} \, \mathbf{f}_i(t),
    \label{eq:langevin}
\end{equation}
where each Cartesian component of $\mathbf{f}_i(t)$ corresponds to a white noise with zero mean and unit variance, 
$\gamma$ denotes the dimensionless viscous damping of the thermal bath and $T$ corresponds to the dimensionless temperature (measured in units of $\epsilon_s/k_B$). The equations of motion were integrated using a Verlet-type stochastic scheme, specifically the G--JF algorithm~\cite{Gronbech-Jensen2013}.}

{Simulations of the system examined in this work exhibit a tendency to become trapped in metastable states due to the inherent frustration arising from the competing length scales in the pair potential. The degeneracy of the ground states combined with finite-size effects further slow convergence to equilibrium, by promoting the formation of arrested phase boundaries and defect structures. To mitigate these issues, we employ a parallel tempering technique (also known as the replica-exchange algorithm), which efficiently explores the configuration space near phase transitions~\cite{Hukushima1996, Sugita1999} and facilitates escape from metastable states at low temperatures by allowing replicas to access higher thermal energies. These simulations are then followed by an {\it annealing} of the replica associated with the lowest temperature down to temperatures approaching zero. This hybrid approach -- together with the use of full inertial Langevin dynamics -- has proven effective in equilibrating low-temperature phases. While parallel tempering alleviates major equilibration barriers, the final annealing step suppresses thermal fluctuations, thereby revealing the underlying ground-state pattern.}

\begin{figure}[ht!]
    \centering
    \resizebox{!}{.40\textwidth}{\begingroup
  \makeatletter
  \providecommand\color[2][]{\GenericError{(gnuplot) \space\space\space\@spaces}{Package color not loaded in conjunction with
      terminal option `colourtext'}{See the gnuplot documentation for explanation.}{Either use 'blacktext' in gnuplot or load the package
      color.sty in LaTeX.}\renewcommand\color[2][]{}}\providecommand\includegraphics[2][]{\GenericError{(gnuplot) \space\space\space\@spaces}{Package graphicx or graphics not loaded}{See the gnuplot documentation for explanation.}{The gnuplot epslatex terminal needs graphicx.sty or graphics.sty.}\renewcommand\includegraphics[2][]{}}\providecommand\rotatebox[2]{#2}\@ifundefined{ifGPcolor}{\newif\ifGPcolor
    \GPcolortrue
  }{}\@ifundefined{ifGPblacktext}{\newif\ifGPblacktext
    \GPblacktexttrue
  }{}\let\gplgaddtomacro\g@addto@macro
\gdef\gplbacktext{}\gdef\gplfronttext{}\makeatother
  \ifGPblacktext
\def\colorrgb#1{}\def\colorgray#1{}\else
\ifGPcolor
      \def\colorrgb#1{\color[rgb]{#1}}\def\colorgray#1{\color[gray]{#1}}\expandafter\def\csname LTw\endcsname{\color{white}}\expandafter\def\csname LTb\endcsname{\color{black}}\expandafter\def\csname LTa\endcsname{\color{black}}\expandafter\def\csname LT0\endcsname{\color[rgb]{1,0,0}}\expandafter\def\csname LT1\endcsname{\color[rgb]{0,1,0}}\expandafter\def\csname LT2\endcsname{\color[rgb]{0,0,1}}\expandafter\def\csname LT3\endcsname{\color[rgb]{1,0,1}}\expandafter\def\csname LT4\endcsname{\color[rgb]{0,1,1}}\expandafter\def\csname LT5\endcsname{\color[rgb]{1,1,0}}\expandafter\def\csname LT6\endcsname{\color[rgb]{0,0,0}}\expandafter\def\csname LT7\endcsname{\color[rgb]{1,0.3,0}}\expandafter\def\csname LT8\endcsname{\color[rgb]{0.5,0.5,0.5}}\else
\def\colorrgb#1{\color{black}}\def\colorgray#1{\color[gray]{#1}}\expandafter\def\csname LTw\endcsname{\color{white}}\expandafter\def\csname LTb\endcsname{\color{black}}\expandafter\def\csname LTa\endcsname{\color{black}}\expandafter\def\csname LT0\endcsname{\color{black}}\expandafter\def\csname LT1\endcsname{\color{black}}\expandafter\def\csname LT2\endcsname{\color{black}}\expandafter\def\csname LT3\endcsname{\color{black}}\expandafter\def\csname LT4\endcsname{\color{black}}\expandafter\def\csname LT5\endcsname{\color{black}}\expandafter\def\csname LT6\endcsname{\color{black}}\expandafter\def\csname LT7\endcsname{\color{black}}\expandafter\def\csname LT8\endcsname{\color{black}}\fi
  \fi
    \setlength{\unitlength}{0.0500bp}\ifx\gptboxheight\undefined \newlength{\gptboxheight}\newlength{\gptboxwidth}\newsavebox{\gptboxtext}\fi \setlength{\fboxrule}{0.5pt}\setlength{\fboxsep}{1pt}\definecolor{tbcol}{rgb}{1,1,1}\begin{picture}(4320.00,4320.00)\gplgaddtomacro\gplbacktext{\csname LTb\endcsname \put(150,4019){\makebox(0,0)[l]{\strut{}a)}}}\gplgaddtomacro\gplfronttext{\csname LTb\endcsname \put(2129,4019){\makebox(0,0){\strut{}$\rho=1.34$\quad$\ell_C=3.00$\quad$T=0.0016$}}}\gplgaddtomacro\gplbacktext{}\gplgaddtomacro\gplfronttext{}\gplbacktext
    \put(0,0){\includegraphics[width={216.00bp},height={216.00bp}]{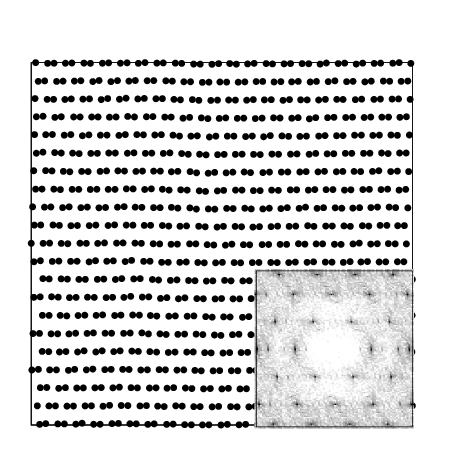}}\gplfronttext
  \end{picture}\endgroup
 }
    \resizebox{!}{.40\textwidth}{\begingroup
  \makeatletter
  \providecommand\color[2][]{\GenericError{(gnuplot) \space\space\space\@spaces}{Package color not loaded in conjunction with
      terminal option `colourtext'}{See the gnuplot documentation for explanation.}{Either use 'blacktext' in gnuplot or load the package
      color.sty in LaTeX.}\renewcommand\color[2][]{}}\providecommand\includegraphics[2][]{\GenericError{(gnuplot) \space\space\space\@spaces}{Package graphicx or graphics not loaded}{See the gnuplot documentation for explanation.}{The gnuplot epslatex terminal needs graphicx.sty or graphics.sty.}\renewcommand\includegraphics[2][]{}}\providecommand\rotatebox[2]{#2}\@ifundefined{ifGPcolor}{\newif\ifGPcolor
    \GPcolortrue
  }{}\@ifundefined{ifGPblacktext}{\newif\ifGPblacktext
    \GPblacktexttrue
  }{}\let\gplgaddtomacro\g@addto@macro
\gdef\gplbacktext{}\gdef\gplfronttext{}\makeatother
  \ifGPblacktext
\def\colorrgb#1{}\def\colorgray#1{}\else
\ifGPcolor
      \def\colorrgb#1{\color[rgb]{#1}}\def\colorgray#1{\color[gray]{#1}}\expandafter\def\csname LTw\endcsname{\color{white}}\expandafter\def\csname LTb\endcsname{\color{black}}\expandafter\def\csname LTa\endcsname{\color{black}}\expandafter\def\csname LT0\endcsname{\color[rgb]{1,0,0}}\expandafter\def\csname LT1\endcsname{\color[rgb]{0,1,0}}\expandafter\def\csname LT2\endcsname{\color[rgb]{0,0,1}}\expandafter\def\csname LT3\endcsname{\color[rgb]{1,0,1}}\expandafter\def\csname LT4\endcsname{\color[rgb]{0,1,1}}\expandafter\def\csname LT5\endcsname{\color[rgb]{1,1,0}}\expandafter\def\csname LT6\endcsname{\color[rgb]{0,0,0}}\expandafter\def\csname LT7\endcsname{\color[rgb]{1,0.3,0}}\expandafter\def\csname LT8\endcsname{\color[rgb]{0.5,0.5,0.5}}\else
\def\colorrgb#1{\color{black}}\def\colorgray#1{\color[gray]{#1}}\expandafter\def\csname LTw\endcsname{\color{white}}\expandafter\def\csname LTb\endcsname{\color{black}}\expandafter\def\csname LTa\endcsname{\color{black}}\expandafter\def\csname LT0\endcsname{\color{black}}\expandafter\def\csname LT1\endcsname{\color{black}}\expandafter\def\csname LT2\endcsname{\color{black}}\expandafter\def\csname LT3\endcsname{\color{black}}\expandafter\def\csname LT4\endcsname{\color{black}}\expandafter\def\csname LT5\endcsname{\color{black}}\expandafter\def\csname LT6\endcsname{\color{black}}\expandafter\def\csname LT7\endcsname{\color{black}}\expandafter\def\csname LT8\endcsname{\color{black}}\fi
  \fi
    \setlength{\unitlength}{0.0500bp}\ifx\gptboxheight\undefined \newlength{\gptboxheight}\newlength{\gptboxwidth}\newsavebox{\gptboxtext}\fi \setlength{\fboxrule}{0.5pt}\setlength{\fboxsep}{1pt}\definecolor{tbcol}{rgb}{1,1,1}\begin{picture}(4320.00,4320.00)\gplgaddtomacro\gplbacktext{\csname LTb\endcsname \put(150,4019){\makebox(0,0)[l]{\strut{}b)}}}\gplgaddtomacro\gplfronttext{\csname LTb\endcsname \put(2129,4019){\makebox(0,0){\strut{}$\rho=1.15$\quad$\ell_C=3.00$\quad$T=0.0015$}}}\gplgaddtomacro\gplbacktext{}\gplgaddtomacro\gplfronttext{}\gplbacktext
    \put(0,0){\includegraphics[width={216.00bp},height={216.00bp}]{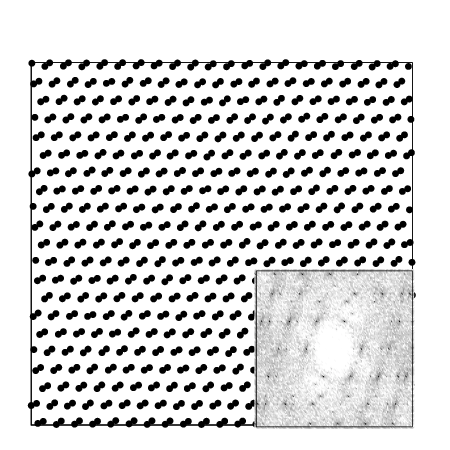}}\gplfronttext
  \end{picture}\endgroup
 }
    
    \resizebox{!}{.40\textwidth}{\begingroup
  \makeatletter
  \providecommand\color[2][]{\GenericError{(gnuplot) \space\space\space\@spaces}{Package color not loaded in conjunction with
      terminal option `colourtext'}{See the gnuplot documentation for explanation.}{Either use 'blacktext' in gnuplot or load the package
      color.sty in LaTeX.}\renewcommand\color[2][]{}}\providecommand\includegraphics[2][]{\GenericError{(gnuplot) \space\space\space\@spaces}{Package graphicx or graphics not loaded}{See the gnuplot documentation for explanation.}{The gnuplot epslatex terminal needs graphicx.sty or graphics.sty.}\renewcommand\includegraphics[2][]{}}\providecommand\rotatebox[2]{#2}\@ifundefined{ifGPcolor}{\newif\ifGPcolor
    \GPcolortrue
  }{}\@ifundefined{ifGPblacktext}{\newif\ifGPblacktext
    \GPblacktexttrue
  }{}\let\gplgaddtomacro\g@addto@macro
\gdef\gplbacktext{}\gdef\gplfronttext{}\makeatother
  \ifGPblacktext
\def\colorrgb#1{}\def\colorgray#1{}\else
\ifGPcolor
      \def\colorrgb#1{\color[rgb]{#1}}\def\colorgray#1{\color[gray]{#1}}\expandafter\def\csname LTw\endcsname{\color{white}}\expandafter\def\csname LTb\endcsname{\color{black}}\expandafter\def\csname LTa\endcsname{\color{black}}\expandafter\def\csname LT0\endcsname{\color[rgb]{1,0,0}}\expandafter\def\csname LT1\endcsname{\color[rgb]{0,1,0}}\expandafter\def\csname LT2\endcsname{\color[rgb]{0,0,1}}\expandafter\def\csname LT3\endcsname{\color[rgb]{1,0,1}}\expandafter\def\csname LT4\endcsname{\color[rgb]{0,1,1}}\expandafter\def\csname LT5\endcsname{\color[rgb]{1,1,0}}\expandafter\def\csname LT6\endcsname{\color[rgb]{0,0,0}}\expandafter\def\csname LT7\endcsname{\color[rgb]{1,0.3,0}}\expandafter\def\csname LT8\endcsname{\color[rgb]{0.5,0.5,0.5}}\else
\def\colorrgb#1{\color{black}}\def\colorgray#1{\color[gray]{#1}}\expandafter\def\csname LTw\endcsname{\color{white}}\expandafter\def\csname LTb\endcsname{\color{black}}\expandafter\def\csname LTa\endcsname{\color{black}}\expandafter\def\csname LT0\endcsname{\color{black}}\expandafter\def\csname LT1\endcsname{\color{black}}\expandafter\def\csname LT2\endcsname{\color{black}}\expandafter\def\csname LT3\endcsname{\color{black}}\expandafter\def\csname LT4\endcsname{\color{black}}\expandafter\def\csname LT5\endcsname{\color{black}}\expandafter\def\csname LT6\endcsname{\color{black}}\expandafter\def\csname LT7\endcsname{\color{black}}\expandafter\def\csname LT8\endcsname{\color{black}}\fi
  \fi
    \setlength{\unitlength}{0.0500bp}\ifx\gptboxheight\undefined \newlength{\gptboxheight}\newlength{\gptboxwidth}\newsavebox{\gptboxtext}\fi \setlength{\fboxrule}{0.5pt}\setlength{\fboxsep}{1pt}\definecolor{tbcol}{rgb}{1,1,1}\begin{picture}(4320.00,4320.00)\gplgaddtomacro\gplbacktext{\csname LTb\endcsname \put(150,4019){\makebox(0,0)[l]{\strut{}c)}}}\gplgaddtomacro\gplfronttext{\csname LTb\endcsname \put(2129,4019){\makebox(0,0){\strut{}$\rho=1.95$\quad$\ell_C=3.50$\quad$T=0.004$}}}\gplgaddtomacro\gplbacktext{}\gplgaddtomacro\gplfronttext{}\gplbacktext
    \put(0,0){\includegraphics[width={216.00bp},height={216.00bp}]{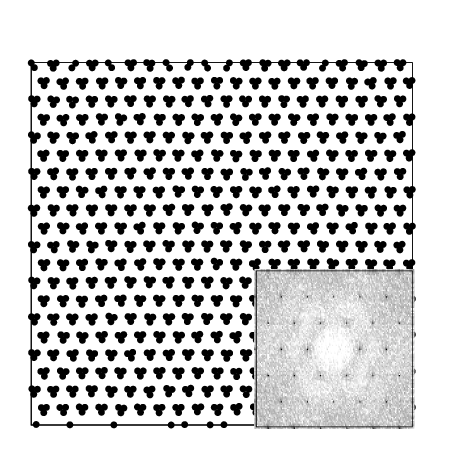}}\gplfronttext
  \end{picture}\endgroup
 }
    \resizebox{!}{.40\textwidth}{\begingroup
  \makeatletter
  \providecommand\color[2][]{\GenericError{(gnuplot) \space\space\space\@spaces}{Package color not loaded in conjunction with
      terminal option `colourtext'}{See the gnuplot documentation for explanation.}{Either use 'blacktext' in gnuplot or load the package
      color.sty in LaTeX.}\renewcommand\color[2][]{}}\providecommand\includegraphics[2][]{\GenericError{(gnuplot) \space\space\space\@spaces}{Package graphicx or graphics not loaded}{See the gnuplot documentation for explanation.}{The gnuplot epslatex terminal needs graphicx.sty or graphics.sty.}\renewcommand\includegraphics[2][]{}}\providecommand\rotatebox[2]{#2}\@ifundefined{ifGPcolor}{\newif\ifGPcolor
    \GPcolortrue
  }{}\@ifundefined{ifGPblacktext}{\newif\ifGPblacktext
    \GPblacktexttrue
  }{}\let\gplgaddtomacro\g@addto@macro
\gdef\gplbacktext{}\gdef\gplfronttext{}\makeatother
  \ifGPblacktext
\def\colorrgb#1{}\def\colorgray#1{}\else
\ifGPcolor
      \def\colorrgb#1{\color[rgb]{#1}}\def\colorgray#1{\color[gray]{#1}}\expandafter\def\csname LTw\endcsname{\color{white}}\expandafter\def\csname LTb\endcsname{\color{black}}\expandafter\def\csname LTa\endcsname{\color{black}}\expandafter\def\csname LT0\endcsname{\color[rgb]{1,0,0}}\expandafter\def\csname LT1\endcsname{\color[rgb]{0,1,0}}\expandafter\def\csname LT2\endcsname{\color[rgb]{0,0,1}}\expandafter\def\csname LT3\endcsname{\color[rgb]{1,0,1}}\expandafter\def\csname LT4\endcsname{\color[rgb]{0,1,1}}\expandafter\def\csname LT5\endcsname{\color[rgb]{1,1,0}}\expandafter\def\csname LT6\endcsname{\color[rgb]{0,0,0}}\expandafter\def\csname LT7\endcsname{\color[rgb]{1,0.3,0}}\expandafter\def\csname LT8\endcsname{\color[rgb]{0.5,0.5,0.5}}\else
\def\colorrgb#1{\color{black}}\def\colorgray#1{\color[gray]{#1}}\expandafter\def\csname LTw\endcsname{\color{white}}\expandafter\def\csname LTb\endcsname{\color{black}}\expandafter\def\csname LTa\endcsname{\color{black}}\expandafter\def\csname LT0\endcsname{\color{black}}\expandafter\def\csname LT1\endcsname{\color{black}}\expandafter\def\csname LT2\endcsname{\color{black}}\expandafter\def\csname LT3\endcsname{\color{black}}\expandafter\def\csname LT4\endcsname{\color{black}}\expandafter\def\csname LT5\endcsname{\color{black}}\expandafter\def\csname LT6\endcsname{\color{black}}\expandafter\def\csname LT7\endcsname{\color{black}}\expandafter\def\csname LT8\endcsname{\color{black}}\fi
  \fi
    \setlength{\unitlength}{0.0500bp}\ifx\gptboxheight\undefined \newlength{\gptboxheight}\newlength{\gptboxwidth}\newsavebox{\gptboxtext}\fi \setlength{\fboxrule}{0.5pt}\setlength{\fboxsep}{1pt}\definecolor{tbcol}{rgb}{1,1,1}\begin{picture}(4320.00,4320.00)\gplgaddtomacro\gplbacktext{\csname LTb\endcsname \put(150,4019){\makebox(0,0)[l]{\strut{}d)}}}\gplgaddtomacro\gplfronttext{\csname LTb\endcsname \put(2129,4019){\makebox(0,0){\strut{}$\rho=1.65$\quad$\ell_C=3.50$\quad$T=0.0018$}}}\gplgaddtomacro\gplbacktext{}\gplgaddtomacro\gplfronttext{}\gplbacktext
    \put(0,0){\includegraphics[width={216.00bp},height={216.00bp}]{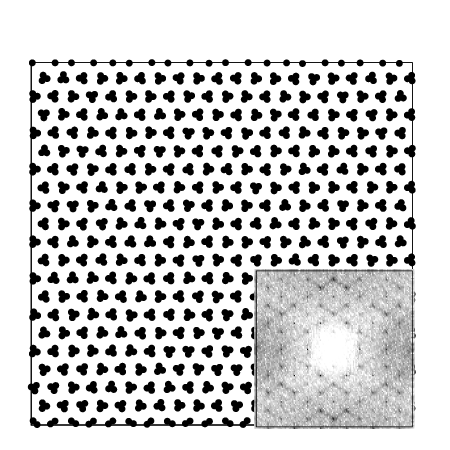}}\gplfronttext
  \end{picture}\endgroup
 }
\caption{{
    Cluster crystal low temperature molecular dynamics configurations corresponding to dimer (a,b) and trimer (c, d) phases, for systems with $N=900$ and $N=1200$ particles, respectively.    
    The corresponding values of the density, relative strength of the hard-core interaction and temperature are provided above each configuration.    
    The associated structure factors of the configurations are shown in the insets.}}
    \label{fig:simu1}
\end{figure}

{We have performed simulations with a number of particles ranging $N$ from 400 to 1200 particles using $m=1$, $\gamma=0.05$ and $dt=0.005$ in eq.~(\ref{eq:langevin}). For the parallel tempering we have used approximately $n_T\simeq 100$ temperatures, distributed according to the Hukushima-Nemoto algorithm~\cite{Hukushima1996} such that the swapping ratio between neighboring replicas is almost constant and independent of the temperature values -- the highest temperature values ranging from $0.06$ to $0.12$, depending on the phase diagram region. Figure~\ref{fig:simu1} shows representative configurations obtained from these simulations in the case of cluster crystal phases with occupation number 2 and 3. As an initial survey, we performed a broad sweep of the phase diagram using increments $\Delta \rho = 0.05$ and $\Delta \ell_C = 0.05$, enabling us to identify the overall structure of the phase diagram with moderate resolution. These final configurations from the simulations served as an inspiration for the construction of the several variational ansatz developed to investigate the ground-state phase diagram in this system.}

{Nevertheless, as will become clear shortly, more general ansatz than those observed in the simulations were constructed, so the simulations have not limited our results in any aspect. While one could attempt to sketch a phase diagram based solely on numerical simulations, this strategy has notable drawbacks: as discussed earlier, simulations may settle into metastable configurations, which makes it difficult to obtain a reliable comparison between the energies of different ground-states. In addition, simulation results alone do not produce a reliable characterization of the nature of phase transitions and identification of possible coexistence regions — features that become accessible when the energy of each ground-state candidate is analytically determined, as it will become clear throughout the present work in the following sections. Conversely, the simulations are particularly valuable for exposing quasiperiodic phases, for which constructing a suitable ansatz is far from trivial. The quasicrystalline phases obtained from simulations will be examined in more detail at the end of Sect.~\ref{results}.}

\subsection{From generic to specific ansatz}

\begin{figure}[ht!]
  \centering
  \includesvg[width=0.3\linewidth]{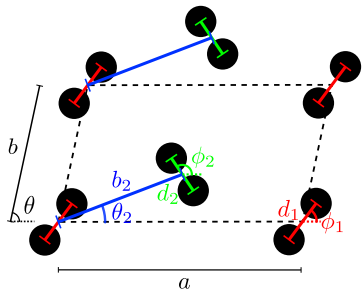}
  \caption{Schematic representation of the general ansatz employed to describe cluster 2 ground-states. The variational parameters considered are highlighted in the figure. Parameters $a$, $b$ and $\theta$ characterize the lattice of clusters whereas $d_1$ and $\phi_1$ and $d_2$ and $\phi_2$ describe the structure of the dimers on each sub-lattice. Finally, the parameters $b_2$ and $\theta_2$ give us the relative position of the second sub-lattice respect to first one.}
  \label{fig:general_ansatz}
\end{figure}

The simulations results suggested that, in the low-density region of interest $\rho<2.5$, corresponding to short-wavelength ground states, the cluster phases have occupation numbers between 2 and 4. To explore as many possible cluster ground states as feasible with energy minimization, we began by considering a generic ansatz, illustrated for cluster 2 in Figure~\ref{fig:general_ansatz}. Rather than assuming a triangular lattice of clusters, we allowed a general angle $\theta$ between primitive translation vectors and variable lattice parameters along each primitive direction, described by $a$ along the x-axis and $b$ along the $\theta$ direction. The distance between particles within a cluster is denoted $d_1$, oriented along angle $\phi_1$. Additionally, we allowed the existence of a second sub-lattice, identical to the original one, but displaced  by a distance $b_2$ at angle $\theta_2$, with its own cluster size and orientation described by $d_2$ and $\phi_2$. Similar general ansatz were constructed for clusters with occupation numbers 3 and 4.

As with the molecular dynamics stage, the aim of the numerical minimization based on these general ansatz was exploratory. The parameters $a$, $b$ and $\theta$ are not constrained, and in principle can represent any of the five two-dimensional Bravais lattices (oblique, rectangular, centered rectangular, triangular, and square). These degrees of freedom, for example, enable a hexagonal lattice to adapt to significant cluster size distortions. The same general ansatz can also describe non-Bravais lattices, such as those identified in previous works~\cite{Rossini2018, Mambretti2021}, where cluster 2 phases exhibit dimers with non-uniform orientations alternating along one lattice direction.

{
We can evaluate the energy per particle of the system by summing the potential energy corresponding to all pairs of particles in our system:
\begin{equation}\label{eq:generic_ansatz}
  \frac{E}{N} = \frac{1}{2 n s}\left(\sum_{l,m}^\infty\sum_{q,q'}^n\sum_{r,r'}^s V(|\textbf{R}_{lm}+\textbf{s}_r-\textbf{s}_{r'}+\textbf{u}_{q,r}-\textbf{u}_{q',r'}|)\right),
\end{equation}
where $n$ is the cluster size and $s$ is the number of sub-lattices (in our case $s=2$). Moreover, the set $\{\textbf{R}_{lm}\}$ and $\{\textbf{u}_{q,r}\}$ represent the positions of the center of the clusters in a sub-lattice, while $\textbf{s}_r$ represent the position of the corresponding sub-lattice respect to the first sub-lattice (so $\textbf{s}_1 = \textbf{0}$). Naturally, the indexes $l$ and $m$ serve to identify a specific lattice vector once we specify the corresponding lattice vector basis. As an example, to reproduce the configuration in Figure \ref{fig:general_ansatz}, we should set:
\begin{equation}
\begin{aligned}
\textbf{R}_{lm}&\equiv l\textbf{a}+m\textbf{b},\\
\textbf{s}_2&\equiv\textbf{b}_2,
\end{aligned}
\end{equation}
where the set $\{\textbf{a},\textbf{b}\}$ and $\textbf{b}_2$ represent the oblique vector basis of the lattice and the relative position of the second sub-lattice, respectively (see Fig. \ref{fig:general_ansatz}).} 

The lattice positions given by these vectors will be then decorated by the cluster structure, which can be described by a finite size triangular lattice. For the cluster 2 geometry, for instance, the positions can be parametrized as:
\begin{equation}
\textbf{u}_{q,r}=\pm\frac{d_{r}}{2}\left\{\cos\phi_{r}, \sin\phi_{r}\right\},\\
\end{equation}
where the $r$ index identifies the corresponding sub-lattice (1 or 2), whereas $d_r$ and $\phi_r$ represent the corresponding dimer size and orientation.

For the numerical minimization, we used the Sequential Quadratic Programming (SQP) solver for nonlinearly constrained problems from the ALGLIB library, version 3.16 \cite{alglib}. The optional use of constraints helped avoid misidentification of phases. One issue with the general ansatz is that it cannot always sharply resolve phase boundaries or transitions, since it can adapt to different phases. For example, the dimer phase ansatz can also describe a triangular crystal. Another challenge is the large number of variational parameters, which requires testing different initial conditions to deal with the many metastable states. Nevertheless, combined with the molecular dynamics guidance, this approach allowed us to systematically identify the phases present in the system.

\begin{figure}[ht!]
    \centering
    \includesvg[width=0.95\linewidth]{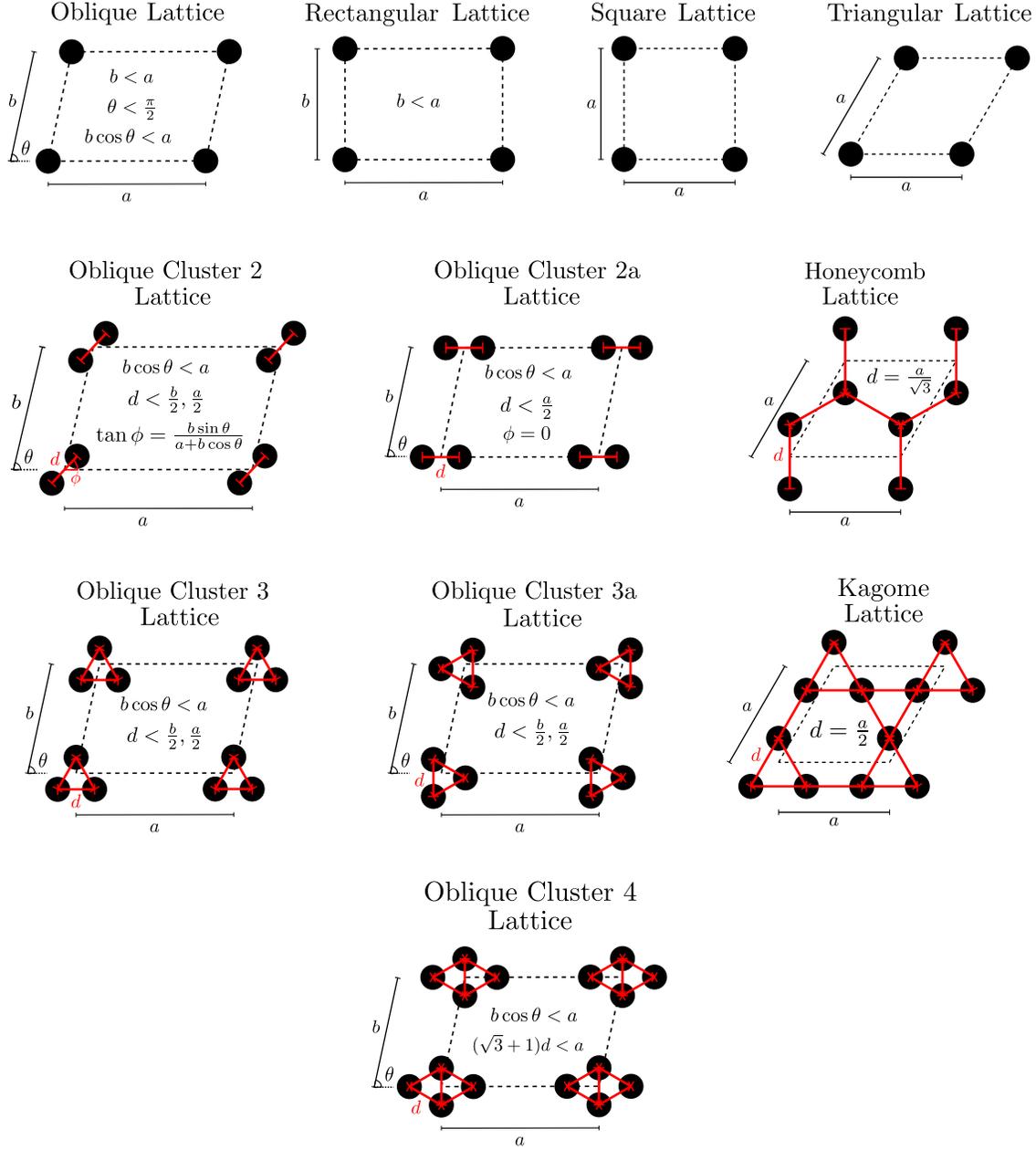}
    \caption{Geometric representation of the different ansatz considered for each phase. The occupation number increases from the top to the bottom row. Each figure corresponds to a possible solution of our ground state phase diagram. The variational parameters, together with the main constraints on them, are presented for each configuration.}
    \label{fig:specific_ansatz}
\end{figure}

We identified traditional Bravais lattice phases with occupancy number $n=1$, such as {\it triangular} and {\it square} lattices, as well as single-particle {\it rectangular} and {\it oblique} lattices, corresponding in mesophase terminology to stripe phases with unit width. Their specific ansatz are shown in the first row of Figure~\ref{fig:specific_ansatz}.

We also found two cluster 2 (dimer) phases, both in general assuming the form of oblique lattices. In the first, dimers are oriented along the line equidistant to neighboring dimers — described by the angle $\phi=\arctan( b\sin\theta/(a+b\cos\theta) )$ — referred to here as the {\it oblique cluster 2 lattice}, or simply Cluster 2. In the second, the dimers align along a primitive direction of the oblique lattice, referred to as the {\it oblique cluster 2-aligned lattice}, or just Cluster 2a. Additionally, we identified a {\it honeycomb} lattice, corresponding to a cluster 2 ``inverse micelle'' or ``hole''-type phase. The specific ansatz for these are illustrated in the second row of Fig.~\ref{fig:specific_ansatz}.

For cluster 3, we identified two phases: one with homogeneous trimer orientation, called the {\it oblique cluster 3 lattice} (or Cluster 3), and another with alternating trimer orientation along one primitive direction of the oblique lattice, designated the {\it oblique cluster 3-alternated lattice} (or Cluster 3a). Larger trimers can form a {\it kagome lattice}, corresponding to a cluster 3 hole-type phase. The third row of Fig.\ref{fig:specific_ansatz} shows these, while the fourth row presents the ansatz for the {\it oblique cluster 4 lattice} (or Cluster 4 for simplicity), which is also capable to describe mesophases associated with cluster 4, such as stripes and holes, as the cluster distance decreases and clusters merge. For simplicity, we group all such cluster 4 mesophases under the general description of the oblique cluster 4 lattice phase.

{This set of 11 different specific ansatz, represented in Figure \ref{fig:specific_ansatz}, constitutes the basis for constructing the zero-temperature phase diagram of the system. Each was used to
evaluate the energy as:
\begin{equation}\label{eq:specific_ansatz}
  \frac{E}{N} = \frac{1}{2 n}\left(\sum_{l,m}^\infty\sum_{q,q'}^n V(|\textbf{R}_{lm}+\textbf{u}_{qlm}-\textbf{u}_{q'00}|)\right),
\end{equation}
where $\textbf{R}_{lm}\equiv l\textbf{a}+m\textbf{b}$ and $\textbf{u}_{(q+1)lm}=\{d\cos(\phi_{lm}+q\pi/3), d\sin(\phi_{lm}+q\pi/3)\}$, for $q\geq 0$. The site dependent phases $\phi_{lm}$ are taken taken as a single variational parameter for all Cluster 3 phases except for the case of Cluster 3a phase in which we have two different equivalent sub-lattices with opposite cluster orientation. Here the index convention is the same as in Eq.(\ref{eq:generic_ansatz}). Note that in general, especially for larger clusters, it may be necessary to modify the $\textbf{u}_q$, however, up to cluster 4, our parametrization will accurately describe cluster geometries. Despite the fact that Cluster 3a phase have two different sub-lattices the validity of Eq.~(\ref{eq:specific_ansatz}) in this case stems from the fact that both sub-lattices forming the crystal are equivalent by a lattice reflection using $\textbf{a}$ as symmetry axis.}

Furthermore, we should notice that at a given density $\rho$ the lattice parameters are not independent. In general, the density of particles can be evaluated as
\begin{equation}
    \rho = {n_{c}}/{A}, 
\end{equation}
where $n_{c}$ is the number of particles inside the basic cell of the pattern having an area $A$. For example, for the oblique lattice the basic cell area is given as $A=a b \sin\theta$, which lead us to the conclusion: 
\begin{equation}
b=\frac{n_{c}}{a \rho \sin\theta}.
\end{equation}

Taking into account this kind of constraint, we have as independent variational parameters $a$, $\theta$ and $d$ for all cluster phases, $a$ and $\theta$ for the oblique phase and just $a$ for the rectangular phase. On the other hand, the triangular, square, honeycomb and kagome phases are fully determined by the density, without any minimization. In order to enforce each ansatz to represent a single type of configuration we should impose additional constraints to the variational parameters as presented in {Figure~\ref{fig:specific_ansatz}}. {For each ansatz, we minimized its energy with respect to its variational parameters at each point in the ($\rho, \ell_C$) plane. To efficiently construct the phase diagram, for each $\ell_C$ we implemented a sweeping protocol in density, where the optimized parameters for each ansatz at a given $\rho$ served as the initial condition for the same ansatz in the next point $\rho+\Delta\rho$.}

\section{Results}
\label{results}

\subsection{Phase diagram}
The systematic determination of the ground-state phase diagram was carried out over a wide range of densities $\rho\in[0.7,2.5]$ and values of the hard-core strength parameter $\ell_C \in [0.5, 4.0]$. Figure~3 summarizes the phase behavior of the system under these parameters. Several important observations emerge from this diagram.

\begin{figure}[ht!]
    \centering
    \begingroup
  \makeatletter
  \providecommand\color[2][]{\GenericError{(gnuplot) \space\space\space\@spaces}{Package color not loaded in conjunction with
      terminal option `colourtext'}{See the gnuplot documentation for explanation.}{Either use 'blacktext' in gnuplot or load the package
      color.sty in LaTeX.}\renewcommand\color[2][]{}}\providecommand\includegraphics[2][]{\GenericError{(gnuplot) \space\space\space\@spaces}{Package graphicx or graphics not loaded}{See the gnuplot documentation for explanation.}{The gnuplot epslatex terminal needs graphicx.sty or graphics.sty.}\renewcommand\includegraphics[2][]{}}\providecommand\rotatebox[2]{#2}\@ifundefined{ifGPcolor}{\newif\ifGPcolor
    \GPcolorfalse
  }{}\@ifundefined{ifGPblacktext}{\newif\ifGPblacktext
    \GPblacktexttrue
  }{}\let\gplgaddtomacro\g@addto@macro
\gdef\gplbacktext{}\gdef\gplfronttext{}\makeatother
  \ifGPblacktext
\def\colorrgb#1{}\def\colorgray#1{}\else
\ifGPcolor
      \def\colorrgb#1{\color[rgb]{#1}}\def\colorgray#1{\color[gray]{#1}}\expandafter\def\csname LTw\endcsname{\color{white}}\expandafter\def\csname LTb\endcsname{\color{black}}\expandafter\def\csname LTa\endcsname{\color{black}}\expandafter\def\csname LT0\endcsname{\color[rgb]{1,0,0}}\expandafter\def\csname LT1\endcsname{\color[rgb]{0,1,0}}\expandafter\def\csname LT2\endcsname{\color[rgb]{0,0,1}}\expandafter\def\csname LT3\endcsname{\color[rgb]{1,0,1}}\expandafter\def\csname LT4\endcsname{\color[rgb]{0,1,1}}\expandafter\def\csname LT5\endcsname{\color[rgb]{1,1,0}}\expandafter\def\csname LT6\endcsname{\color[rgb]{0,0,0}}\expandafter\def\csname LT7\endcsname{\color[rgb]{1,0.3,0}}\expandafter\def\csname LT8\endcsname{\color[rgb]{0.5,0.5,0.5}}\else
\def\colorrgb#1{\color{black}}\def\colorgray#1{\color[gray]{#1}}\expandafter\def\csname LTw\endcsname{\color{white}}\expandafter\def\csname LTb\endcsname{\color{black}}\expandafter\def\csname LTa\endcsname{\color{black}}\expandafter\def\csname LT0\endcsname{\color{black}}\expandafter\def\csname LT1\endcsname{\color{black}}\expandafter\def\csname LT2\endcsname{\color{black}}\expandafter\def\csname LT3\endcsname{\color{black}}\expandafter\def\csname LT4\endcsname{\color{black}}\expandafter\def\csname LT5\endcsname{\color{black}}\expandafter\def\csname LT6\endcsname{\color{black}}\expandafter\def\csname LT7\endcsname{\color{black}}\expandafter\def\csname LT8\endcsname{\color{black}}\fi
  \fi
    \setlength{\unitlength}{0.0500bp}\ifx\gptboxheight\undefined \newlength{\gptboxheight}\newlength{\gptboxwidth}\newsavebox{\gptboxtext}\fi \setlength{\fboxrule}{0.5pt}\setlength{\fboxsep}{1pt}\begin{picture}(6048.00,4030.00)\gplgaddtomacro\gplbacktext{\csname LTb\endcsname \put(740,3829){\makebox(0,0)[r]{\strut{}0.5}}\put(740,3373){\makebox(0,0)[r]{\strut{}1.0}}\put(740,2918){\makebox(0,0)[r]{\strut{}1.5}}\put(740,2462){\makebox(0,0)[r]{\strut{}2.0}}\put(740,2007){\makebox(0,0)[r]{\strut{}2.5}}\put(740,1551){\makebox(0,0)[r]{\strut{}3.0}}\put(740,1096){\makebox(0,0)[r]{\strut{}3.5}}\put(740,640){\makebox(0,0)[r]{\strut{}4.0}}\put(1255,440){\makebox(0,0){\strut{}0.9}}\put(1847,440){\makebox(0,0){\strut{}1.2}}\put(2439,440){\makebox(0,0){\strut{}1.5}}\put(3031,440){\makebox(0,0){\strut{}1.8}}\put(3623,440){\makebox(0,0){\strut{}2.1}}\put(4215,440){\makebox(0,0){\strut{}2.4}}}\gplgaddtomacro\gplfronttext{\csname LTb\endcsname \put(130,2234){\rotatebox{-270}{\makebox(0,0){\strut{}$\ell_C$}}}\put(2636,140){\makebox(0,0){\strut{}$\rho$}}\csname LTb\endcsname \put(740,3829){\makebox(0,0)[r]{\strut{}0.5}}\put(740,3373){\makebox(0,0)[r]{\strut{}1.0}}\put(740,2918){\makebox(0,0)[r]{\strut{}1.5}}\put(740,2462){\makebox(0,0)[r]{\strut{}2.0}}\put(740,2007){\makebox(0,0)[r]{\strut{}2.5}}\put(740,1551){\makebox(0,0)[r]{\strut{}3.0}}\put(740,1096){\makebox(0,0)[r]{\strut{}3.5}}\put(740,640){\makebox(0,0)[r]{\strut{}4.0}}\put(1255,440){\makebox(0,0){\strut{}0.9}}\put(1847,440){\makebox(0,0){\strut{}1.2}}\put(2439,440){\makebox(0,0){\strut{}1.5}}\put(3031,440){\makebox(0,0){\strut{}1.8}}\put(3623,440){\makebox(0,0){\strut{}2.1}}\put(4215,440){\makebox(0,0){\strut{}2.4}}\put(4798,784){\makebox(0,0)[l]{\strut{}Cluster 4}}\put(4798,1074){\makebox(0,0)[l]{\strut{}Cluster 3a}}\put(4798,1364){\makebox(0,0)[l]{\strut{}Cluster 3}}\put(4798,1654){\makebox(0,0)[l]{\strut{}Cluster 2a}}\put(4798,1944){\makebox(0,0)[l]{\strut{}Cluster 2}}\put(4798,2234){\makebox(0,0)[l]{\strut{}Kagome}}\put(4798,2524){\makebox(0,0)[l]{\strut{}Honeycomb}}\put(4798,2814){\makebox(0,0)[l]{\strut{}Oblique}}\put(4798,3104){\makebox(0,0)[l]{\strut{}Rectangular}}\put(4798,3394){\makebox(0,0)[l]{\strut{}Square}}\put(4798,3684){\makebox(0,0)[l]{\strut{}Triangular}}}\gplbacktext
    \put(0,0){\includegraphics{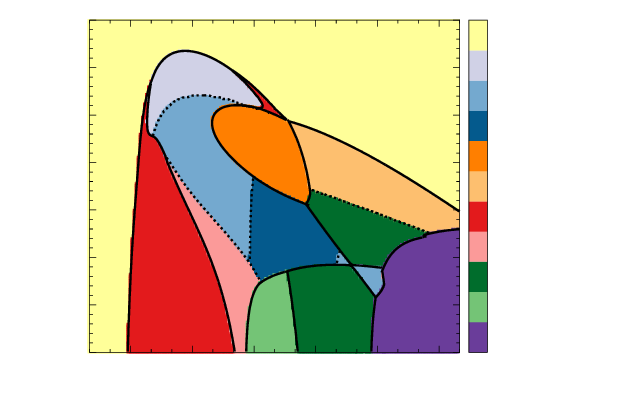}}\gplfronttext
  \end{picture}\endgroup
     \caption{Ground state phase diagram for model presented in Eq.~(\ref{pot}) using the density $\rho$ and relative strength of the hard-core interaction $\ell_C = -\log C$ as running parameters. The phase boundaries are represented with full lines in the case of first-order transitions, and by dashed lines in the case of continuous transitions.}
    \label{fig:diag}
\end{figure}

To begin with, the overall structure of the phase diagram confirms that the addition of a hard-core repulsion to the GEM-4 ultra-soft interaction strongly enriches the phase behavior, suppressing the otherwise {contiguous} sequence of cluster occupancies characteristic of the pure GEM-4 models~\cite{Likos2007, Neuhaus_2011, Mello_2023}. Indeed, looking at the smallest values of $C$ explored (highest values of $\ell_C$), i.e., for weakest hard-core repulsion, one observes a sequence of cluster phases with occupancy numbers that grows with increasing density. This is expected from the known behavior of cluster crystal phases in systems with pure ultra-soft potentials, but with two remarkable differences. The first being the presence of more than one phase associated with a given occupation number. The second is that, as it will be discussed in detail shortly, the oblique lattice formed by the center of the clusters in some cases is closer to a triangular lattice than in others, depending on the occupancy number. 

Naturally, for higher values of $C$ or lower values of $\ell_C$, the cluster size turns comparable to the lattice spacing between clusters, and clusters crystals phases give space for new single-particle phases, including square, rectangular, oblique, honeycomb and kagome phases. This demonstrates how the hard-core term disrupts particle overlap and favors unusual lattice structures in the context of ultra-soft pair potentials. 

An interesting feature of the obtained phase diagram is the presence of both first-order (solid lines) and continuous (dashed lines) phase transitions. The first-order transitions generally appear between phases that are incommensurate with each other. These phases have distinct local symmetries or incompatible coordination motifs, preventing a continuous deformation from one to the other. As a consequence, the system exhibits abrupt changes in the particle distribution — typically involving changes in the local cluster occupancy or orientational order of the clusters — signaling a discontinuous jump in the absolute energy minimum configuration. A clear example is the transition from Cluster~2 to Cluster~3 phases, where the number of particles per cluster changes discretely.

By contrast, the dashed lines indicate continuous transitions, where the symmetry or local motif of one phase can be smoothly transformed into another by adjusting lattice parameters. For example, the Oblique lattice phase can continuously approach a Rectangular lattice phase as its lattice angle $\theta$ tends toward $\pi/2$, and the Rectangular lattice phase can evolve further into a Square lattice phase if its aspect ratio reaches unity. Similarly, a Cluster~2a phase can be continuously tuned into a Rectangular lattice phase as its internal distance $d$ grows. These pathways highlight the role of commensurability between phases, which defines whether a smooth transition is possible.

The presence of the Cluster~2a and Cluster~3a phases deserves emphasis. These phases appear as intermediates states between the more stable Cluster 2 and Cluster 3 phases. In particular, Cluster~2a exhibits a dimer structure oriented along one of the main directions of the lattice in a configuration that is reminiscent of a 
single particle width stripe phase, as is the case of the Oblique and Rectangular phases. In Cluster~3a, a more complex intra-cluster structure develop, suggesting a frustration between maximizing intra-cluster packing and optimizing the inter-cluster repulsion. Their relatively narrow stability windows reflect a delicate balance between these competing interactions, showing how the hard-core potential creates new local frustration absent in the purely ultra-soft case. In overall, the phase diagram illustrates how the introduction of a competing length scale through a hard-core term creates an expanded palette of mesophases in two dimensions, breaking the classical scenario of purely cluster-crystal states.

\subsection{Ground-state configurations}

In Figure~4, we show detailed representations of the ground-state configurations obtained by minimizing the energy within the specific ansatz defined previously. These snapshots illustrate the microscopic ordering of each phase and allow us to interpret their structural features in more depth.

\begin{figure}[ht!]
  \centering
  \resizebox{!}{.24\textwidth}{\begingroup
  \makeatletter
\@ifundefined{ifGPcolor}{\newif\ifGPcolor
    \GPcolorfalse
  }{}\@ifundefined{ifGPblacktext}{\newif\ifGPblacktext
    \GPblacktexttrue
  }{}\let\gplgaddtomacro\g@addto@macro
\gdef\gplbacktext{}\gdef\gplfronttext{}\makeatother
  \ifGPblacktext
\def\colorrgb#1{}\def\colorgray#1{}\else
\ifGPcolor
      \def\colorrgb#1{\color[rgb]{#1}}\def\colorgray#1{\color[gray]{#1}}\expandafter\def\csname LTw\endcsname{\color{white}}\expandafter\def\csname LTb\endcsname{\color{black}}\expandafter\def\csname LTa\endcsname{\color{black}}\expandafter\def\csname LT0\endcsname{\color[rgb]{1,0,0}}\expandafter\def\csname LT1\endcsname{\color[rgb]{0,1,0}}\expandafter\def\csname LT2\endcsname{\color[rgb]{0,0,1}}\expandafter\def\csname LT3\endcsname{\color[rgb]{1,0,1}}\expandafter\def\csname LT4\endcsname{\color[rgb]{0,1,1}}\expandafter\def\csname LT5\endcsname{\color[rgb]{1,1,0}}\expandafter\def\csname LT6\endcsname{\color[rgb]{0,0,0}}\expandafter\def\csname LT7\endcsname{\color[rgb]{1,0.3,0}}\expandafter\def\csname LT8\endcsname{\color[rgb]{0.5,0.5,0.5}}\else
\def\colorrgb#1{\color{black}}\def\colorgray#1{\color[gray]{#1}}\expandafter\def\csname LTw\endcsname{\color{white}}\expandafter\def\csname LTb\endcsname{\color{black}}\expandafter\def\csname LTa\endcsname{\color{black}}\expandafter\def\csname LT0\endcsname{\color{black}}\expandafter\def\csname LT1\endcsname{\color{black}}\expandafter\def\csname LT2\endcsname{\color{black}}\expandafter\def\csname LT3\endcsname{\color{black}}\expandafter\def\csname LT4\endcsname{\color{black}}\expandafter\def\csname LT5\endcsname{\color{black}}\expandafter\def\csname LT6\endcsname{\color{black}}\expandafter\def\csname LT7\endcsname{\color{black}}\expandafter\def\csname LT8\endcsname{\color{black}}\fi
  \fi
    \setlength{\unitlength}{0.0500bp}\ifx\gptboxheight\undefined \newlength{\gptboxheight}\newlength{\gptboxwidth}\newsavebox{\gptboxtext}\fi \setlength{\fboxrule}{0.5pt}\setlength{\fboxsep}{1pt}\begin{picture}(2880.00,2880.00)\gplgaddtomacro\gplbacktext{\csname LTb\endcsname \put(198,2494){\makebox(0,0)[l]{\strut{}a)}}}\gplgaddtomacro\gplfronttext{\csname LTb\endcsname \put(1409,2579){\makebox(0,0){\strut{}Cluster 2}}\put(1409,2379){\makebox(0,0){\strut{}$\rho=1.15 \quad \ell_C=3.01$}}}\gplbacktext
    \put(0,0){\includegraphics{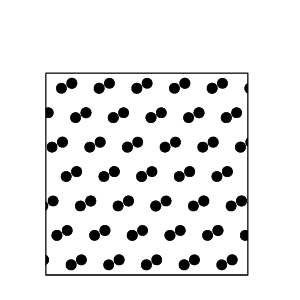}}\gplfronttext
  \end{picture}\endgroup
 \begingroup
  \makeatletter
  \providecommand\color[2][]{\GenericError{(gnuplot) \space\space\space\@spaces}{Package color not loaded in conjunction with
      terminal option `colourtext'}{See the gnuplot documentation for explanation.}{Either use 'blacktext' in gnuplot or load the package
      color.sty in LaTeX.}\renewcommand\color[2][]{}}\providecommand\includegraphics[2][]{\GenericError{(gnuplot) \space\space\space\@spaces}{Package graphicx or graphics not loaded}{See the gnuplot documentation for explanation.}{The gnuplot epslatex terminal needs graphicx.sty or graphics.sty.}\renewcommand\includegraphics[2][]{}}\providecommand\rotatebox[2]{#2}\@ifundefined{ifGPcolor}{\newif\ifGPcolor
    \GPcolorfalse
  }{}\@ifundefined{ifGPblacktext}{\newif\ifGPblacktext
    \GPblacktexttrue
  }{}\let\gplgaddtomacro\g@addto@macro
\gdef\gplbacktext{}\gdef\gplfronttext{}\makeatother
  \ifGPblacktext
\def\colorrgb#1{}\def\colorgray#1{}\else
\ifGPcolor
      \def\colorrgb#1{\color[rgb]{#1}}\def\colorgray#1{\color[gray]{#1}}\expandafter\def\csname LTw\endcsname{\color{white}}\expandafter\def\csname LTb\endcsname{\color{black}}\expandafter\def\csname LTa\endcsname{\color{black}}\expandafter\def\csname LT0\endcsname{\color[rgb]{1,0,0}}\expandafter\def\csname LT1\endcsname{\color[rgb]{0,1,0}}\expandafter\def\csname LT2\endcsname{\color[rgb]{0,0,1}}\expandafter\def\csname LT3\endcsname{\color[rgb]{1,0,1}}\expandafter\def\csname LT4\endcsname{\color[rgb]{0,1,1}}\expandafter\def\csname LT5\endcsname{\color[rgb]{1,1,0}}\expandafter\def\csname LT6\endcsname{\color[rgb]{0,0,0}}\expandafter\def\csname LT7\endcsname{\color[rgb]{1,0.3,0}}\expandafter\def\csname LT8\endcsname{\color[rgb]{0.5,0.5,0.5}}\else
\def\colorrgb#1{\color{black}}\def\colorgray#1{\color[gray]{#1}}\expandafter\def\csname LTw\endcsname{\color{white}}\expandafter\def\csname LTb\endcsname{\color{black}}\expandafter\def\csname LTa\endcsname{\color{black}}\expandafter\def\csname LT0\endcsname{\color{black}}\expandafter\def\csname LT1\endcsname{\color{black}}\expandafter\def\csname LT2\endcsname{\color{black}}\expandafter\def\csname LT3\endcsname{\color{black}}\expandafter\def\csname LT4\endcsname{\color{black}}\expandafter\def\csname LT5\endcsname{\color{black}}\expandafter\def\csname LT6\endcsname{\color{black}}\expandafter\def\csname LT7\endcsname{\color{black}}\expandafter\def\csname LT8\endcsname{\color{black}}\fi
  \fi
    \setlength{\unitlength}{0.0500bp}\ifx\gptboxheight\undefined \newlength{\gptboxheight}\newlength{\gptboxwidth}\newsavebox{\gptboxtext}\fi \setlength{\fboxrule}{0.5pt}\setlength{\fboxsep}{1pt}\begin{picture}(2880.00,2880.00)\gplgaddtomacro\gplbacktext{\csname LTb\endcsname \put(198,2494){\makebox(0,0)[l]{\strut{}b)}}}\gplgaddtomacro\gplfronttext{\csname LTb\endcsname \put(1409,2579){\makebox(0,0){\strut{}Cluster 2}}\put(1409,2379){\makebox(0,0){\strut{}$\rho=1.10 \quad \ell_C=2.01$}}}\gplbacktext
    \put(0,0){\includegraphics{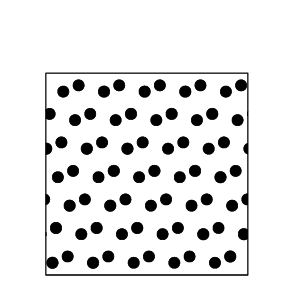}}\gplfronttext
  \end{picture}\endgroup
 \begingroup
  \makeatletter
  \providecommand\color[2][]{\GenericError{(gnuplot) \space\space\space\@spaces}{Package color not loaded in conjunction with
      terminal option `colourtext'}{See the gnuplot documentation for explanation.}{Either use 'blacktext' in gnuplot or load the package
      color.sty in LaTeX.}\renewcommand\color[2][]{}}\providecommand\includegraphics[2][]{\GenericError{(gnuplot) \space\space\space\@spaces}{Package graphicx or graphics not loaded}{See the gnuplot documentation for explanation.}{The gnuplot epslatex terminal needs graphicx.sty or graphics.sty.}\renewcommand\includegraphics[2][]{}}\providecommand\rotatebox[2]{#2}\@ifundefined{ifGPcolor}{\newif\ifGPcolor
    \GPcolorfalse
  }{}\@ifundefined{ifGPblacktext}{\newif\ifGPblacktext
    \GPblacktexttrue
  }{}\let\gplgaddtomacro\g@addto@macro
\gdef\gplbacktext{}\gdef\gplfronttext{}\makeatother
  \ifGPblacktext
\def\colorrgb#1{}\def\colorgray#1{}\else
\ifGPcolor
      \def\colorrgb#1{\color[rgb]{#1}}\def\colorgray#1{\color[gray]{#1}}\expandafter\def\csname LTw\endcsname{\color{white}}\expandafter\def\csname LTb\endcsname{\color{black}}\expandafter\def\csname LTa\endcsname{\color{black}}\expandafter\def\csname LT0\endcsname{\color[rgb]{1,0,0}}\expandafter\def\csname LT1\endcsname{\color[rgb]{0,1,0}}\expandafter\def\csname LT2\endcsname{\color[rgb]{0,0,1}}\expandafter\def\csname LT3\endcsname{\color[rgb]{1,0,1}}\expandafter\def\csname LT4\endcsname{\color[rgb]{0,1,1}}\expandafter\def\csname LT5\endcsname{\color[rgb]{1,1,0}}\expandafter\def\csname LT6\endcsname{\color[rgb]{0,0,0}}\expandafter\def\csname LT7\endcsname{\color[rgb]{1,0.3,0}}\expandafter\def\csname LT8\endcsname{\color[rgb]{0.5,0.5,0.5}}\else
\def\colorrgb#1{\color{black}}\def\colorgray#1{\color[gray]{#1}}\expandafter\def\csname LTw\endcsname{\color{white}}\expandafter\def\csname LTb\endcsname{\color{black}}\expandafter\def\csname LTa\endcsname{\color{black}}\expandafter\def\csname LT0\endcsname{\color{black}}\expandafter\def\csname LT1\endcsname{\color{black}}\expandafter\def\csname LT2\endcsname{\color{black}}\expandafter\def\csname LT3\endcsname{\color{black}}\expandafter\def\csname LT4\endcsname{\color{black}}\expandafter\def\csname LT5\endcsname{\color{black}}\expandafter\def\csname LT6\endcsname{\color{black}}\expandafter\def\csname LT7\endcsname{\color{black}}\expandafter\def\csname LT8\endcsname{\color{black}}\fi
  \fi
    \setlength{\unitlength}{0.0500bp}\ifx\gptboxheight\undefined \newlength{\gptboxheight}\newlength{\gptboxwidth}\newsavebox{\gptboxtext}\fi \setlength{\fboxrule}{0.5pt}\setlength{\fboxsep}{1pt}\begin{picture}(2880.00,2880.00)\gplgaddtomacro\gplbacktext{\csname LTb\endcsname \put(198,2494){\makebox(0,0)[l]{\strut{}c)}}}\gplgaddtomacro\gplfronttext{\csname LTb\endcsname \put(1409,2579){\makebox(0,0){\strut{}Cluster 2a}}\put(1409,2379){\makebox(0,0){\strut{}$\rho=1.35 \quad \ell_C=3.31$}}}\gplbacktext
    \put(0,0){\includegraphics{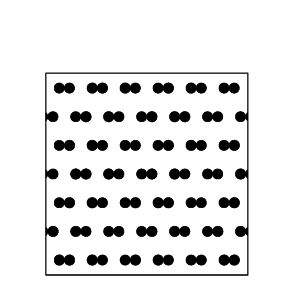}}\gplfronttext
  \end{picture}\endgroup
 \begingroup
  \makeatletter
  \providecommand\color[2][]{\GenericError{(gnuplot) \space\space\space\@spaces}{Package color not loaded in conjunction with
      terminal option `colourtext'}{See the gnuplot documentation for explanation.}{Either use 'blacktext' in gnuplot or load the package
      color.sty in LaTeX.}\renewcommand\color[2][]{}}\providecommand\includegraphics[2][]{\GenericError{(gnuplot) \space\space\space\@spaces}{Package graphicx or graphics not loaded}{See the gnuplot documentation for explanation.}{The gnuplot epslatex terminal needs graphicx.sty or graphics.sty.}\renewcommand\includegraphics[2][]{}}\providecommand\rotatebox[2]{#2}\@ifundefined{ifGPcolor}{\newif\ifGPcolor
    \GPcolorfalse
  }{}\@ifundefined{ifGPblacktext}{\newif\ifGPblacktext
    \GPblacktexttrue
  }{}\let\gplgaddtomacro\g@addto@macro
\gdef\gplbacktext{}\gdef\gplfronttext{}\makeatother
  \ifGPblacktext
\def\colorrgb#1{}\def\colorgray#1{}\else
\ifGPcolor
      \def\colorrgb#1{\color[rgb]{#1}}\def\colorgray#1{\color[gray]{#1}}\expandafter\def\csname LTw\endcsname{\color{white}}\expandafter\def\csname LTb\endcsname{\color{black}}\expandafter\def\csname LTa\endcsname{\color{black}}\expandafter\def\csname LT0\endcsname{\color[rgb]{1,0,0}}\expandafter\def\csname LT1\endcsname{\color[rgb]{0,1,0}}\expandafter\def\csname LT2\endcsname{\color[rgb]{0,0,1}}\expandafter\def\csname LT3\endcsname{\color[rgb]{1,0,1}}\expandafter\def\csname LT4\endcsname{\color[rgb]{0,1,1}}\expandafter\def\csname LT5\endcsname{\color[rgb]{1,1,0}}\expandafter\def\csname LT6\endcsname{\color[rgb]{0,0,0}}\expandafter\def\csname LT7\endcsname{\color[rgb]{1,0.3,0}}\expandafter\def\csname LT8\endcsname{\color[rgb]{0.5,0.5,0.5}}\else
\def\colorrgb#1{\color{black}}\def\colorgray#1{\color[gray]{#1}}\expandafter\def\csname LTw\endcsname{\color{white}}\expandafter\def\csname LTb\endcsname{\color{black}}\expandafter\def\csname LTa\endcsname{\color{black}}\expandafter\def\csname LT0\endcsname{\color{black}}\expandafter\def\csname LT1\endcsname{\color{black}}\expandafter\def\csname LT2\endcsname{\color{black}}\expandafter\def\csname LT3\endcsname{\color{black}}\expandafter\def\csname LT4\endcsname{\color{black}}\expandafter\def\csname LT5\endcsname{\color{black}}\expandafter\def\csname LT6\endcsname{\color{black}}\expandafter\def\csname LT7\endcsname{\color{black}}\expandafter\def\csname LT8\endcsname{\color{black}}\fi
  \fi
    \setlength{\unitlength}{0.0500bp}\ifx\gptboxheight\undefined \newlength{\gptboxheight}\newlength{\gptboxwidth}\newsavebox{\gptboxtext}\fi \setlength{\fboxrule}{0.5pt}\setlength{\fboxsep}{1pt}\begin{picture}(2880.00,2880.00)\gplgaddtomacro\gplbacktext{\csname LTb\endcsname \put(198,2494){\makebox(0,0)[l]{\strut{}d)}}}\gplgaddtomacro\gplfronttext{\csname LTb\endcsname \put(1409,2579){\makebox(0,0){\strut{}Honeycomb}}\put(1409,2379){\makebox(0,0){\strut{}$\rho=1.50 \quad \ell_C=1.71$}}}\gplbacktext
    \put(0,0){\includegraphics{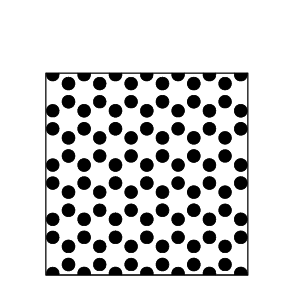}}\gplfronttext
  \end{picture}\endgroup
 }
  \resizebox{!}{.24\textwidth}{\begingroup
  \makeatletter
  \providecommand\color[2][]{\GenericError{(gnuplot) \space\space\space\@spaces}{Package color not loaded in conjunction with
      terminal option `colourtext'}{See the gnuplot documentation for explanation.}{Either use 'blacktext' in gnuplot or load the package
      color.sty in LaTeX.}\renewcommand\color[2][]{}}\providecommand\includegraphics[2][]{\GenericError{(gnuplot) \space\space\space\@spaces}{Package graphicx or graphics not loaded}{See the gnuplot documentation for explanation.}{The gnuplot epslatex terminal needs graphicx.sty or graphics.sty.}\renewcommand\includegraphics[2][]{}}\providecommand\rotatebox[2]{#2}\@ifundefined{ifGPcolor}{\newif\ifGPcolor
    \GPcolorfalse
  }{}\@ifundefined{ifGPblacktext}{\newif\ifGPblacktext
    \GPblacktexttrue
  }{}\let\gplgaddtomacro\g@addto@macro
\gdef\gplbacktext{}\gdef\gplfronttext{}\makeatother
  \ifGPblacktext
\def\colorrgb#1{}\def\colorgray#1{}\else
\ifGPcolor
      \def\colorrgb#1{\color[rgb]{#1}}\def\colorgray#1{\color[gray]{#1}}\expandafter\def\csname LTw\endcsname{\color{white}}\expandafter\def\csname LTb\endcsname{\color{black}}\expandafter\def\csname LTa\endcsname{\color{black}}\expandafter\def\csname LT0\endcsname{\color[rgb]{1,0,0}}\expandafter\def\csname LT1\endcsname{\color[rgb]{0,1,0}}\expandafter\def\csname LT2\endcsname{\color[rgb]{0,0,1}}\expandafter\def\csname LT3\endcsname{\color[rgb]{1,0,1}}\expandafter\def\csname LT4\endcsname{\color[rgb]{0,1,1}}\expandafter\def\csname LT5\endcsname{\color[rgb]{1,1,0}}\expandafter\def\csname LT6\endcsname{\color[rgb]{0,0,0}}\expandafter\def\csname LT7\endcsname{\color[rgb]{1,0.3,0}}\expandafter\def\csname LT8\endcsname{\color[rgb]{0.5,0.5,0.5}}\else
\def\colorrgb#1{\color{black}}\def\colorgray#1{\color[gray]{#1}}\expandafter\def\csname LTw\endcsname{\color{white}}\expandafter\def\csname LTb\endcsname{\color{black}}\expandafter\def\csname LTa\endcsname{\color{black}}\expandafter\def\csname LT0\endcsname{\color{black}}\expandafter\def\csname LT1\endcsname{\color{black}}\expandafter\def\csname LT2\endcsname{\color{black}}\expandafter\def\csname LT3\endcsname{\color{black}}\expandafter\def\csname LT4\endcsname{\color{black}}\expandafter\def\csname LT5\endcsname{\color{black}}\expandafter\def\csname LT6\endcsname{\color{black}}\expandafter\def\csname LT7\endcsname{\color{black}}\expandafter\def\csname LT8\endcsname{\color{black}}\fi
  \fi
    \setlength{\unitlength}{0.0500bp}\ifx\gptboxheight\undefined \newlength{\gptboxheight}\newlength{\gptboxwidth}\newsavebox{\gptboxtext}\fi \setlength{\fboxrule}{0.5pt}\setlength{\fboxsep}{1pt}\begin{picture}(2880.00,2880.00)\gplgaddtomacro\gplbacktext{\csname LTb\endcsname \put(198,2494){\makebox(0,0)[l]{\strut{}e)}}}\gplgaddtomacro\gplfronttext{\csname LTb\endcsname \put(1409,2579){\makebox(0,0){\strut{}Cluster 2a}}\put(1409,2379){\makebox(0,0){\strut{}$\rho=1.25 \quad \ell_C=2.51$}}}\gplbacktext
    \put(0,0){\includegraphics{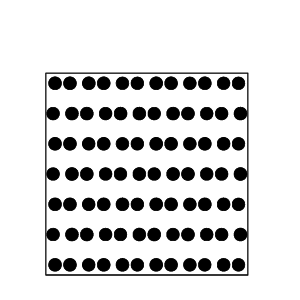}}\gplfronttext
  \end{picture}\endgroup
 \begingroup
  \makeatletter
  \providecommand\color[2][]{\GenericError{(gnuplot) \space\space\space\@spaces}{Package color not loaded in conjunction with
      terminal option `colourtext'}{See the gnuplot documentation for explanation.}{Either use 'blacktext' in gnuplot or load the package
      color.sty in LaTeX.}\renewcommand\color[2][]{}}\providecommand\includegraphics[2][]{\GenericError{(gnuplot) \space\space\space\@spaces}{Package graphicx or graphics not loaded}{See the gnuplot documentation for explanation.}{The gnuplot epslatex terminal needs graphicx.sty or graphics.sty.}\renewcommand\includegraphics[2][]{}}\providecommand\rotatebox[2]{#2}\@ifundefined{ifGPcolor}{\newif\ifGPcolor
    \GPcolorfalse
  }{}\@ifundefined{ifGPblacktext}{\newif\ifGPblacktext
    \GPblacktexttrue
  }{}\let\gplgaddtomacro\g@addto@macro
\gdef\gplbacktext{}\gdef\gplfronttext{}\makeatother
  \ifGPblacktext
\def\colorrgb#1{}\def\colorgray#1{}\else
\ifGPcolor
      \def\colorrgb#1{\color[rgb]{#1}}\def\colorgray#1{\color[gray]{#1}}\expandafter\def\csname LTw\endcsname{\color{white}}\expandafter\def\csname LTb\endcsname{\color{black}}\expandafter\def\csname LTa\endcsname{\color{black}}\expandafter\def\csname LT0\endcsname{\color[rgb]{1,0,0}}\expandafter\def\csname LT1\endcsname{\color[rgb]{0,1,0}}\expandafter\def\csname LT2\endcsname{\color[rgb]{0,0,1}}\expandafter\def\csname LT3\endcsname{\color[rgb]{1,0,1}}\expandafter\def\csname LT4\endcsname{\color[rgb]{0,1,1}}\expandafter\def\csname LT5\endcsname{\color[rgb]{1,1,0}}\expandafter\def\csname LT6\endcsname{\color[rgb]{0,0,0}}\expandafter\def\csname LT7\endcsname{\color[rgb]{1,0.3,0}}\expandafter\def\csname LT8\endcsname{\color[rgb]{0.5,0.5,0.5}}\else
\def\colorrgb#1{\color{black}}\def\colorgray#1{\color[gray]{#1}}\expandafter\def\csname LTw\endcsname{\color{white}}\expandafter\def\csname LTb\endcsname{\color{black}}\expandafter\def\csname LTa\endcsname{\color{black}}\expandafter\def\csname LT0\endcsname{\color{black}}\expandafter\def\csname LT1\endcsname{\color{black}}\expandafter\def\csname LT2\endcsname{\color{black}}\expandafter\def\csname LT3\endcsname{\color{black}}\expandafter\def\csname LT4\endcsname{\color{black}}\expandafter\def\csname LT5\endcsname{\color{black}}\expandafter\def\csname LT6\endcsname{\color{black}}\expandafter\def\csname LT7\endcsname{\color{black}}\expandafter\def\csname LT8\endcsname{\color{black}}\fi
  \fi
    \setlength{\unitlength}{0.0500bp}\ifx\gptboxheight\undefined \newlength{\gptboxheight}\newlength{\gptboxwidth}\newsavebox{\gptboxtext}\fi \setlength{\fboxrule}{0.5pt}\setlength{\fboxsep}{1pt}\begin{picture}(2880.00,2880.00)\gplgaddtomacro\gplbacktext{\csname LTb\endcsname \put(198,2494){\makebox(0,0)[l]{\strut{}f)}}}\gplgaddtomacro\gplfronttext{\csname LTb\endcsname \put(1409,2579){\makebox(0,0){\strut{}Rectangular}}\put(1409,2379){\makebox(0,0){\strut{}$\rho=1.40 \quad \ell_C=2.51$}}}\gplbacktext
    \put(0,0){\includegraphics{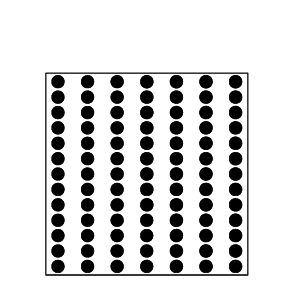}}\gplfronttext
  \end{picture}\endgroup
 \begingroup
  \makeatletter
  \providecommand\color[2][]{\GenericError{(gnuplot) \space\space\space\@spaces}{Package color not loaded in conjunction with
      terminal option `colourtext'}{See the gnuplot documentation for explanation.}{Either use 'blacktext' in gnuplot or load the package
      color.sty in LaTeX.}\renewcommand\color[2][]{}}\providecommand\includegraphics[2][]{\GenericError{(gnuplot) \space\space\space\@spaces}{Package graphicx or graphics not loaded}{See the gnuplot documentation for explanation.}{The gnuplot epslatex terminal needs graphicx.sty or graphics.sty.}\renewcommand\includegraphics[2][]{}}\providecommand\rotatebox[2]{#2}\@ifundefined{ifGPcolor}{\newif\ifGPcolor
    \GPcolorfalse
  }{}\@ifundefined{ifGPblacktext}{\newif\ifGPblacktext
    \GPblacktexttrue
  }{}\let\gplgaddtomacro\g@addto@macro
\gdef\gplbacktext{}\gdef\gplfronttext{}\makeatother
  \ifGPblacktext
\def\colorrgb#1{}\def\colorgray#1{}\else
\ifGPcolor
      \def\colorrgb#1{\color[rgb]{#1}}\def\colorgray#1{\color[gray]{#1}}\expandafter\def\csname LTw\endcsname{\color{white}}\expandafter\def\csname LTb\endcsname{\color{black}}\expandafter\def\csname LTa\endcsname{\color{black}}\expandafter\def\csname LT0\endcsname{\color[rgb]{1,0,0}}\expandafter\def\csname LT1\endcsname{\color[rgb]{0,1,0}}\expandafter\def\csname LT2\endcsname{\color[rgb]{0,0,1}}\expandafter\def\csname LT3\endcsname{\color[rgb]{1,0,1}}\expandafter\def\csname LT4\endcsname{\color[rgb]{0,1,1}}\expandafter\def\csname LT5\endcsname{\color[rgb]{1,1,0}}\expandafter\def\csname LT6\endcsname{\color[rgb]{0,0,0}}\expandafter\def\csname LT7\endcsname{\color[rgb]{1,0.3,0}}\expandafter\def\csname LT8\endcsname{\color[rgb]{0.5,0.5,0.5}}\else
\def\colorrgb#1{\color{black}}\def\colorgray#1{\color[gray]{#1}}\expandafter\def\csname LTw\endcsname{\color{white}}\expandafter\def\csname LTb\endcsname{\color{black}}\expandafter\def\csname LTa\endcsname{\color{black}}\expandafter\def\csname LT0\endcsname{\color{black}}\expandafter\def\csname LT1\endcsname{\color{black}}\expandafter\def\csname LT2\endcsname{\color{black}}\expandafter\def\csname LT3\endcsname{\color{black}}\expandafter\def\csname LT4\endcsname{\color{black}}\expandafter\def\csname LT5\endcsname{\color{black}}\expandafter\def\csname LT6\endcsname{\color{black}}\expandafter\def\csname LT7\endcsname{\color{black}}\expandafter\def\csname LT8\endcsname{\color{black}}\fi
  \fi
    \setlength{\unitlength}{0.0500bp}\ifx\gptboxheight\undefined \newlength{\gptboxheight}\newlength{\gptboxwidth}\newsavebox{\gptboxtext}\fi \setlength{\fboxrule}{0.5pt}\setlength{\fboxsep}{1pt}\begin{picture}(2880.00,2880.00)\gplgaddtomacro\gplbacktext{\csname LTb\endcsname \put(198,2494){\makebox(0,0)[l]{\strut{}g)}}}\gplgaddtomacro\gplfronttext{\csname LTb\endcsname \put(1409,2579){\makebox(0,0){\strut{}Oblique}}\put(1409,2379){\makebox(0,0){\strut{}$\rho=1.60 \quad \ell_C=2.51$}}}\gplbacktext
    \put(0,0){\includegraphics{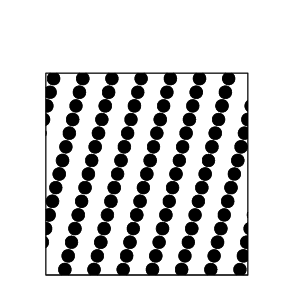}}\gplfronttext
  \end{picture}\endgroup
 \begingroup
  \makeatletter
  \providecommand\color[2][]{\GenericError{(gnuplot) \space\space\space\@spaces}{Package color not loaded in conjunction with
      terminal option `colourtext'}{See the gnuplot documentation for explanation.}{Either use 'blacktext' in gnuplot or load the package
      color.sty in LaTeX.}\renewcommand\color[2][]{}}\providecommand\includegraphics[2][]{\GenericError{(gnuplot) \space\space\space\@spaces}{Package graphicx or graphics not loaded}{See the gnuplot documentation for explanation.}{The gnuplot epslatex terminal needs graphicx.sty or graphics.sty.}\renewcommand\includegraphics[2][]{}}\providecommand\rotatebox[2]{#2}\@ifundefined{ifGPcolor}{\newif\ifGPcolor
    \GPcolorfalse
  }{}\@ifundefined{ifGPblacktext}{\newif\ifGPblacktext
    \GPblacktexttrue
  }{}\let\gplgaddtomacro\g@addto@macro
\gdef\gplbacktext{}\gdef\gplfronttext{}\makeatother
  \ifGPblacktext
\def\colorrgb#1{}\def\colorgray#1{}\else
\ifGPcolor
      \def\colorrgb#1{\color[rgb]{#1}}\def\colorgray#1{\color[gray]{#1}}\expandafter\def\csname LTw\endcsname{\color{white}}\expandafter\def\csname LTb\endcsname{\color{black}}\expandafter\def\csname LTa\endcsname{\color{black}}\expandafter\def\csname LT0\endcsname{\color[rgb]{1,0,0}}\expandafter\def\csname LT1\endcsname{\color[rgb]{0,1,0}}\expandafter\def\csname LT2\endcsname{\color[rgb]{0,0,1}}\expandafter\def\csname LT3\endcsname{\color[rgb]{1,0,1}}\expandafter\def\csname LT4\endcsname{\color[rgb]{0,1,1}}\expandafter\def\csname LT5\endcsname{\color[rgb]{1,1,0}}\expandafter\def\csname LT6\endcsname{\color[rgb]{0,0,0}}\expandafter\def\csname LT7\endcsname{\color[rgb]{1,0.3,0}}\expandafter\def\csname LT8\endcsname{\color[rgb]{0.5,0.5,0.5}}\else
\def\colorrgb#1{\color{black}}\def\colorgray#1{\color[gray]{#1}}\expandafter\def\csname LTw\endcsname{\color{white}}\expandafter\def\csname LTb\endcsname{\color{black}}\expandafter\def\csname LTa\endcsname{\color{black}}\expandafter\def\csname LT0\endcsname{\color{black}}\expandafter\def\csname LT1\endcsname{\color{black}}\expandafter\def\csname LT2\endcsname{\color{black}}\expandafter\def\csname LT3\endcsname{\color{black}}\expandafter\def\csname LT4\endcsname{\color{black}}\expandafter\def\csname LT5\endcsname{\color{black}}\expandafter\def\csname LT6\endcsname{\color{black}}\expandafter\def\csname LT7\endcsname{\color{black}}\expandafter\def\csname LT8\endcsname{\color{black}}\fi
  \fi
    \setlength{\unitlength}{0.0500bp}\ifx\gptboxheight\undefined \newlength{\gptboxheight}\newlength{\gptboxwidth}\newsavebox{\gptboxtext}\fi \setlength{\fboxrule}{0.5pt}\setlength{\fboxsep}{1pt}\begin{picture}(2880.00,2880.00)\gplgaddtomacro\gplbacktext{\csname LTb\endcsname \put(198,2494){\makebox(0,0)[l]{\strut{}h)}}}\gplgaddtomacro\gplfronttext{\csname LTb\endcsname \put(1409,2579){\makebox(0,0){\strut{}Kagomé}}\put(1409,2379){\makebox(0,0){\strut{}$\rho=2.10 \quad \ell_C=2.31$}}}\gplbacktext
    \put(0,0){\includegraphics{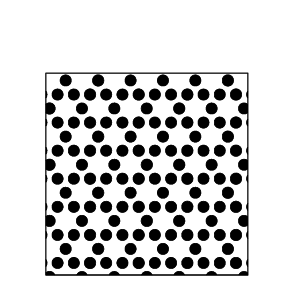}}\gplfronttext
  \end{picture}\endgroup
 }
  \resizebox{!}{.24\textwidth}{\begingroup
  \makeatletter
  \providecommand\color[2][]{\GenericError{(gnuplot) \space\space\space\@spaces}{Package color not loaded in conjunction with
      terminal option `colourtext'}{See the gnuplot documentation for explanation.}{Either use 'blacktext' in gnuplot or load the package
      color.sty in LaTeX.}\renewcommand\color[2][]{}}\providecommand\includegraphics[2][]{\GenericError{(gnuplot) \space\space\space\@spaces}{Package graphicx or graphics not loaded}{See the gnuplot documentation for explanation.}{The gnuplot epslatex terminal needs graphicx.sty or graphics.sty.}\renewcommand\includegraphics[2][]{}}\providecommand\rotatebox[2]{#2}\@ifundefined{ifGPcolor}{\newif\ifGPcolor
    \GPcolorfalse
  }{}\@ifundefined{ifGPblacktext}{\newif\ifGPblacktext
    \GPblacktexttrue
  }{}\let\gplgaddtomacro\g@addto@macro
\gdef\gplbacktext{}\gdef\gplfronttext{}\makeatother
  \ifGPblacktext
\def\colorrgb#1{}\def\colorgray#1{}\else
\ifGPcolor
      \def\colorrgb#1{\color[rgb]{#1}}\def\colorgray#1{\color[gray]{#1}}\expandafter\def\csname LTw\endcsname{\color{white}}\expandafter\def\csname LTb\endcsname{\color{black}}\expandafter\def\csname LTa\endcsname{\color{black}}\expandafter\def\csname LT0\endcsname{\color[rgb]{1,0,0}}\expandafter\def\csname LT1\endcsname{\color[rgb]{0,1,0}}\expandafter\def\csname LT2\endcsname{\color[rgb]{0,0,1}}\expandafter\def\csname LT3\endcsname{\color[rgb]{1,0,1}}\expandafter\def\csname LT4\endcsname{\color[rgb]{0,1,1}}\expandafter\def\csname LT5\endcsname{\color[rgb]{1,1,0}}\expandafter\def\csname LT6\endcsname{\color[rgb]{0,0,0}}\expandafter\def\csname LT7\endcsname{\color[rgb]{1,0.3,0}}\expandafter\def\csname LT8\endcsname{\color[rgb]{0.5,0.5,0.5}}\else
\def\colorrgb#1{\color{black}}\def\colorgray#1{\color[gray]{#1}}\expandafter\def\csname LTw\endcsname{\color{white}}\expandafter\def\csname LTb\endcsname{\color{black}}\expandafter\def\csname LTa\endcsname{\color{black}}\expandafter\def\csname LT0\endcsname{\color{black}}\expandafter\def\csname LT1\endcsname{\color{black}}\expandafter\def\csname LT2\endcsname{\color{black}}\expandafter\def\csname LT3\endcsname{\color{black}}\expandafter\def\csname LT4\endcsname{\color{black}}\expandafter\def\csname LT5\endcsname{\color{black}}\expandafter\def\csname LT6\endcsname{\color{black}}\expandafter\def\csname LT7\endcsname{\color{black}}\expandafter\def\csname LT8\endcsname{\color{black}}\fi
  \fi
    \setlength{\unitlength}{0.0500bp}\ifx\gptboxheight\undefined \newlength{\gptboxheight}\newlength{\gptboxwidth}\newsavebox{\gptboxtext}\fi \setlength{\fboxrule}{0.5pt}\setlength{\fboxsep}{1pt}\begin{picture}(2880.00,2880.00)\gplgaddtomacro\gplbacktext{\csname LTb\endcsname \put(198,2494){\makebox(0,0)[l]{\strut{}i)}}}\gplgaddtomacro\gplfronttext{\csname LTb\endcsname \put(1409,2579){\makebox(0,0){\strut{}Cluster 3}}\put(1409,2379){\makebox(0,0){\strut{}$\rho=1.80 \quad \ell_C=3.11$}}}\gplbacktext
    \put(0,0){\includegraphics{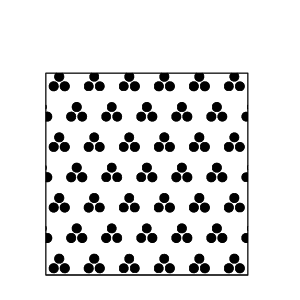}}\gplfronttext
  \end{picture}\endgroup
 \begingroup
  \makeatletter
  \providecommand\color[2][]{\GenericError{(gnuplot) \space\space\space\@spaces}{Package color not loaded in conjunction with
      terminal option `colourtext'}{See the gnuplot documentation for explanation.}{Either use 'blacktext' in gnuplot or load the package
      color.sty in LaTeX.}\renewcommand\color[2][]{}}\providecommand\includegraphics[2][]{\GenericError{(gnuplot) \space\space\space\@spaces}{Package graphicx or graphics not loaded}{See the gnuplot documentation for explanation.}{The gnuplot epslatex terminal needs graphicx.sty or graphics.sty.}\renewcommand\includegraphics[2][]{}}\providecommand\rotatebox[2]{#2}\@ifundefined{ifGPcolor}{\newif\ifGPcolor
    \GPcolorfalse
  }{}\@ifundefined{ifGPblacktext}{\newif\ifGPblacktext
    \GPblacktexttrue
  }{}\let\gplgaddtomacro\g@addto@macro
\gdef\gplbacktext{}\gdef\gplfronttext{}\makeatother
  \ifGPblacktext
\def\colorrgb#1{}\def\colorgray#1{}\else
\ifGPcolor
      \def\colorrgb#1{\color[rgb]{#1}}\def\colorgray#1{\color[gray]{#1}}\expandafter\def\csname LTw\endcsname{\color{white}}\expandafter\def\csname LTb\endcsname{\color{black}}\expandafter\def\csname LTa\endcsname{\color{black}}\expandafter\def\csname LT0\endcsname{\color[rgb]{1,0,0}}\expandafter\def\csname LT1\endcsname{\color[rgb]{0,1,0}}\expandafter\def\csname LT2\endcsname{\color[rgb]{0,0,1}}\expandafter\def\csname LT3\endcsname{\color[rgb]{1,0,1}}\expandafter\def\csname LT4\endcsname{\color[rgb]{0,1,1}}\expandafter\def\csname LT5\endcsname{\color[rgb]{1,1,0}}\expandafter\def\csname LT6\endcsname{\color[rgb]{0,0,0}}\expandafter\def\csname LT7\endcsname{\color[rgb]{1,0.3,0}}\expandafter\def\csname LT8\endcsname{\color[rgb]{0.5,0.5,0.5}}\else
\def\colorrgb#1{\color{black}}\def\colorgray#1{\color[gray]{#1}}\expandafter\def\csname LTw\endcsname{\color{white}}\expandafter\def\csname LTb\endcsname{\color{black}}\expandafter\def\csname LTa\endcsname{\color{black}}\expandafter\def\csname LT0\endcsname{\color{black}}\expandafter\def\csname LT1\endcsname{\color{black}}\expandafter\def\csname LT2\endcsname{\color{black}}\expandafter\def\csname LT3\endcsname{\color{black}}\expandafter\def\csname LT4\endcsname{\color{black}}\expandafter\def\csname LT5\endcsname{\color{black}}\expandafter\def\csname LT6\endcsname{\color{black}}\expandafter\def\csname LT7\endcsname{\color{black}}\expandafter\def\csname LT8\endcsname{\color{black}}\fi
  \fi
    \setlength{\unitlength}{0.0500bp}\ifx\gptboxheight\undefined \newlength{\gptboxheight}\newlength{\gptboxwidth}\newsavebox{\gptboxtext}\fi \setlength{\fboxrule}{0.5pt}\setlength{\fboxsep}{1pt}\begin{picture}(2880.00,2880.00)\gplgaddtomacro\gplbacktext{\csname LTb\endcsname \put(198,2494){\makebox(0,0)[l]{\strut{}j)}}}\gplgaddtomacro\gplfronttext{\csname LTb\endcsname \put(1409,2579){\makebox(0,0){\strut{}Cluster 3}}\put(1409,2379){\makebox(0,0){\strut{}$\rho=2.00 \quad \ell_C=2.71$}}}\gplbacktext
    \put(0,0){\includegraphics{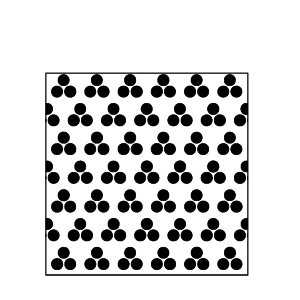}}\gplfronttext
  \end{picture}\endgroup
 \begingroup
  \makeatletter
  \providecommand\color[2][]{\GenericError{(gnuplot) \space\space\space\@spaces}{Package color not loaded in conjunction with
      terminal option `colourtext'}{See the gnuplot documentation for explanation.}{Either use 'blacktext' in gnuplot or load the package
      color.sty in LaTeX.}\renewcommand\color[2][]{}}\providecommand\includegraphics[2][]{\GenericError{(gnuplot) \space\space\space\@spaces}{Package graphicx or graphics not loaded}{See the gnuplot documentation for explanation.}{The gnuplot epslatex terminal needs graphicx.sty or graphics.sty.}\renewcommand\includegraphics[2][]{}}\providecommand\rotatebox[2]{#2}\@ifundefined{ifGPcolor}{\newif\ifGPcolor
    \GPcolorfalse
  }{}\@ifundefined{ifGPblacktext}{\newif\ifGPblacktext
    \GPblacktexttrue
  }{}\let\gplgaddtomacro\g@addto@macro
\gdef\gplbacktext{}\gdef\gplfronttext{}\makeatother
  \ifGPblacktext
\def\colorrgb#1{}\def\colorgray#1{}\else
\ifGPcolor
      \def\colorrgb#1{\color[rgb]{#1}}\def\colorgray#1{\color[gray]{#1}}\expandafter\def\csname LTw\endcsname{\color{white}}\expandafter\def\csname LTb\endcsname{\color{black}}\expandafter\def\csname LTa\endcsname{\color{black}}\expandafter\def\csname LT0\endcsname{\color[rgb]{1,0,0}}\expandafter\def\csname LT1\endcsname{\color[rgb]{0,1,0}}\expandafter\def\csname LT2\endcsname{\color[rgb]{0,0,1}}\expandafter\def\csname LT3\endcsname{\color[rgb]{1,0,1}}\expandafter\def\csname LT4\endcsname{\color[rgb]{0,1,1}}\expandafter\def\csname LT5\endcsname{\color[rgb]{1,1,0}}\expandafter\def\csname LT6\endcsname{\color[rgb]{0,0,0}}\expandafter\def\csname LT7\endcsname{\color[rgb]{1,0.3,0}}\expandafter\def\csname LT8\endcsname{\color[rgb]{0.5,0.5,0.5}}\else
\def\colorrgb#1{\color{black}}\def\colorgray#1{\color[gray]{#1}}\expandafter\def\csname LTw\endcsname{\color{white}}\expandafter\def\csname LTb\endcsname{\color{black}}\expandafter\def\csname LTa\endcsname{\color{black}}\expandafter\def\csname LT0\endcsname{\color{black}}\expandafter\def\csname LT1\endcsname{\color{black}}\expandafter\def\csname LT2\endcsname{\color{black}}\expandafter\def\csname LT3\endcsname{\color{black}}\expandafter\def\csname LT4\endcsname{\color{black}}\expandafter\def\csname LT5\endcsname{\color{black}}\expandafter\def\csname LT6\endcsname{\color{black}}\expandafter\def\csname LT7\endcsname{\color{black}}\expandafter\def\csname LT8\endcsname{\color{black}}\fi
  \fi
    \setlength{\unitlength}{0.0500bp}\ifx\gptboxheight\undefined \newlength{\gptboxheight}\newlength{\gptboxwidth}\newsavebox{\gptboxtext}\fi \setlength{\fboxrule}{0.5pt}\setlength{\fboxsep}{1pt}\begin{picture}(2880.00,2880.00)\gplgaddtomacro\gplbacktext{\csname LTb\endcsname \put(198,2494){\makebox(0,0)[l]{\strut{}k)}}}\gplgaddtomacro\gplfronttext{\csname LTb\endcsname \put(1409,2579){\makebox(0,0){\strut{}Cluster 3a}}\put(1409,2379){\makebox(0,0){\strut{}$\rho=1.65 \quad \ell_C=3.51$}}}\gplbacktext
    \put(0,0){\includegraphics{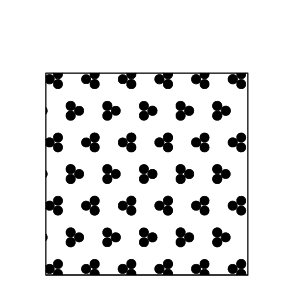}}\gplfronttext
  \end{picture}\endgroup
 \begingroup
  \makeatletter
  \providecommand\color[2][]{\GenericError{(gnuplot) \space\space\space\@spaces}{Package color not loaded in conjunction with
      terminal option `colourtext'}{See the gnuplot documentation for explanation.}{Either use 'blacktext' in gnuplot or load the package
      color.sty in LaTeX.}\renewcommand\color[2][]{}}\providecommand\includegraphics[2][]{\GenericError{(gnuplot) \space\space\space\@spaces}{Package graphicx or graphics not loaded}{See the gnuplot documentation for explanation.}{The gnuplot epslatex terminal needs graphicx.sty or graphics.sty.}\renewcommand\includegraphics[2][]{}}\providecommand\rotatebox[2]{#2}\@ifundefined{ifGPcolor}{\newif\ifGPcolor
    \GPcolorfalse
  }{}\@ifundefined{ifGPblacktext}{\newif\ifGPblacktext
    \GPblacktexttrue
  }{}\let\gplgaddtomacro\g@addto@macro
\gdef\gplbacktext{}\gdef\gplfronttext{}\makeatother
  \ifGPblacktext
\def\colorrgb#1{}\def\colorgray#1{}\else
\ifGPcolor
      \def\colorrgb#1{\color[rgb]{#1}}\def\colorgray#1{\color[gray]{#1}}\expandafter\def\csname LTw\endcsname{\color{white}}\expandafter\def\csname LTb\endcsname{\color{black}}\expandafter\def\csname LTa\endcsname{\color{black}}\expandafter\def\csname LT0\endcsname{\color[rgb]{1,0,0}}\expandafter\def\csname LT1\endcsname{\color[rgb]{0,1,0}}\expandafter\def\csname LT2\endcsname{\color[rgb]{0,0,1}}\expandafter\def\csname LT3\endcsname{\color[rgb]{1,0,1}}\expandafter\def\csname LT4\endcsname{\color[rgb]{0,1,1}}\expandafter\def\csname LT5\endcsname{\color[rgb]{1,1,0}}\expandafter\def\csname LT6\endcsname{\color[rgb]{0,0,0}}\expandafter\def\csname LT7\endcsname{\color[rgb]{1,0.3,0}}\expandafter\def\csname LT8\endcsname{\color[rgb]{0.5,0.5,0.5}}\else
\def\colorrgb#1{\color{black}}\def\colorgray#1{\color[gray]{#1}}\expandafter\def\csname LTw\endcsname{\color{white}}\expandafter\def\csname LTb\endcsname{\color{black}}\expandafter\def\csname LTa\endcsname{\color{black}}\expandafter\def\csname LT0\endcsname{\color{black}}\expandafter\def\csname LT1\endcsname{\color{black}}\expandafter\def\csname LT2\endcsname{\color{black}}\expandafter\def\csname LT3\endcsname{\color{black}}\expandafter\def\csname LT4\endcsname{\color{black}}\expandafter\def\csname LT5\endcsname{\color{black}}\expandafter\def\csname LT6\endcsname{\color{black}}\expandafter\def\csname LT7\endcsname{\color{black}}\expandafter\def\csname LT8\endcsname{\color{black}}\fi
  \fi
    \setlength{\unitlength}{0.0500bp}\ifx\gptboxheight\undefined \newlength{\gptboxheight}\newlength{\gptboxwidth}\newsavebox{\gptboxtext}\fi \setlength{\fboxrule}{0.5pt}\setlength{\fboxsep}{1pt}\begin{picture}(2880.00,2880.00)\gplgaddtomacro\gplbacktext{\csname LTb\endcsname \put(198,2494){\makebox(0,0)[l]{\strut{}l)}}}\gplgaddtomacro\gplfronttext{\csname LTb\endcsname \put(1409,2579){\makebox(0,0){\strut{}Cluster 4}}\put(1409,2379){\makebox(0,0){\strut{}$\rho=2.30 \quad \ell_C=3.51$}}}\gplbacktext
    \put(0,0){\includegraphics{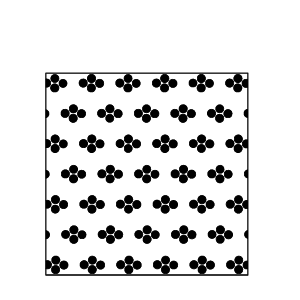}}\gplfronttext
  \end{picture}\endgroup
 }
  \caption{Ground state configurations 
  for selected values of $\rho$ and $\ell_C$ according to the phase diagram in Fig.~\ref{fig:diag}.}
  \label{fig:configs}
\end{figure}

For Cluster~2 and Cluster~2a [panels (a)--(c)], the configurations display a robust tendency to form dimer-like pairs of particles within each cluster. As $\ell_C$ decreases, the distance between particles inside each dimer increases, reflecting the growing influence of the hard-core repulsion. At the same time, the lattice parameter of the inter-cluster structure shrinks, leading to regimes where the intra- and inter-cluster distances become comparable. This is precisely the point where the orientational degree of freedom of the dimers becomes relevant. In the Cluster~2 phase, the dimers are oriented homogeneously along the direction of the midpoint between neighboring dimers, however such orientation varies continuously with $\rho$ and $\ell_C$, introducing a type of orientational order — or nematicity — that is analogous to behaviors seen in dimer-based molecular crystals and frustrated soft matter systems~\cite{Rossini2018}. The possibility of tuning dimer orientations across the lattice marks a nontrivial extension of conventional cluster phases.

The Honeycomb and Kagome configurations [panels (d) and (h)] exhibit phases where the hard-core term dominates and prevents overlap, stabilizing open, high-density structures with high symmetry. These lattices are non-Bravais, but present well-defined local environments with uniform coordination numbers. Known for appearing in conditions of high frustration, such patterns can be interpreted as triangular lattices of holes, in a way that they minimize the repulsive energy, at the expense of bringing closer a number of particles. 
Finally, Cluster~3, Cluster~3a, and Cluster~4 phases [panels (i)--(l)] represent the higher cluster occupancy regimes explored. Their structures reveal triangular intra-cluster order, with the local motif approximately constant as $\rho$ grows. In Cluster~3a, the orientation of the clusters is not uniform across the lattice, alternating the direction of the trimers along one primitive direction. This highlights a frustration scenario where intra- and inter-cluster interactions are not easily satisfied.

\begin{figure}[ht!]
    \centering
    \resizebox{!}{0.35\textwidth}{\begingroup
  \makeatletter
\@ifundefined{ifGPcolor}{\newif\ifGPcolor
    \GPcolorfalse
  }{}\@ifundefined{ifGPblacktext}{\newif\ifGPblacktext
    \GPblacktexttrue
  }{}\let\gplgaddtomacro\g@addto@macro
\gdef\gplbacktext{}\gdef\gplfronttext{}\makeatother
  \ifGPblacktext
\def\colorrgb#1{}\def\colorgray#1{}\else
\ifGPcolor
      \def\colorrgb#1{\color[rgb]{#1}}\def\colorgray#1{\color[gray]{#1}}\expandafter\def\csname LTw\endcsname{\color{white}}\expandafter\def\csname LTb\endcsname{\color{black}}\expandafter\def\csname LTa\endcsname{\color{black}}\expandafter\def\csname LT0\endcsname{\color[rgb]{1,0,0}}\expandafter\def\csname LT1\endcsname{\color[rgb]{0,1,0}}\expandafter\def\csname LT2\endcsname{\color[rgb]{0,0,1}}\expandafter\def\csname LT3\endcsname{\color[rgb]{1,0,1}}\expandafter\def\csname LT4\endcsname{\color[rgb]{0,1,1}}\expandafter\def\csname LT5\endcsname{\color[rgb]{1,1,0}}\expandafter\def\csname LT6\endcsname{\color[rgb]{0,0,0}}\expandafter\def\csname LT7\endcsname{\color[rgb]{1,0.3,0}}\expandafter\def\csname LT8\endcsname{\color[rgb]{0.5,0.5,0.5}}\else
\def\colorrgb#1{\color{black}}\def\colorgray#1{\color[gray]{#1}}\expandafter\def\csname LTw\endcsname{\color{white}}\expandafter\def\csname LTb\endcsname{\color{black}}\expandafter\def\csname LTa\endcsname{\color{black}}\expandafter\def\csname LT0\endcsname{\color{black}}\expandafter\def\csname LT1\endcsname{\color{black}}\expandafter\def\csname LT2\endcsname{\color{black}}\expandafter\def\csname LT3\endcsname{\color{black}}\expandafter\def\csname LT4\endcsname{\color{black}}\expandafter\def\csname LT5\endcsname{\color{black}}\expandafter\def\csname LT6\endcsname{\color{black}}\expandafter\def\csname LT7\endcsname{\color{black}}\expandafter\def\csname LT8\endcsname{\color{black}}\fi
  \fi
    \setlength{\unitlength}{0.0500bp}\ifx\gptboxheight\undefined \newlength{\gptboxheight}\newlength{\gptboxwidth}\newsavebox{\gptboxtext}\fi \setlength{\fboxrule}{0.5pt}\setlength{\fboxsep}{1pt}\begin{picture}(5472.00,4030.00)\gplgaddtomacro\gplbacktext{\csname LTb\endcsname \put(860,640){\makebox(0,0)[r]{\strut{}40.0}}\put(860,1365){\makebox(0,0)[r]{\strut{}45.0}}\put(860,2090){\makebox(0,0)[r]{\strut{}50.0}}\put(860,2814){\makebox(0,0)[r]{\strut{}55.0}}\put(860,3539){\makebox(0,0)[r]{\strut{}60.0}}\put(4735,440){\makebox(0,0){\strut{}2.0}}\put(3797,440){\makebox(0,0){\strut{}2.5}}\put(2858,440){\makebox(0,0){\strut{}3.0}}\put(1919,440){\makebox(0,0){\strut{}3.5}}\put(980,440){\makebox(0,0){\strut{}4.0}}\put(41,3829){\makebox(0,0)[l]{\strut{}a)}}}\gplgaddtomacro\gplfronttext{\csname LTb\endcsname \put(130,2234){\rotatebox{-270}{\makebox(0,0){\strut{}{\large $\theta$}}}}\put(3045,140){\makebox(0,0){\strut{}$\ell_C$}}\put(2951,1853){\makebox(0,0){\strut{}$\rho$}}\csname LTb\endcsname \put(3140,1603){\makebox(0,0)[r]{\strut{}(Cluster 2) $0.97$}}\csname LTb\endcsname \put(3140,1403){\makebox(0,0)[r]{\strut{}(Cluster 2) $1.10$}}\csname LTb\endcsname \put(3140,1203){\makebox(0,0)[r]{\strut{}(Cluster 2a) $1.42$}}\csname LTb\endcsname \put(3140,1003){\makebox(0,0)[r]{\strut{}(Cluster 3a) $1.50$}}\csname LTb\endcsname \put(3140,803){\makebox(0,0)[r]{\strut{}(Cluster 3) $1.80$}}}\gplbacktext
    \put(0,0){\includegraphics{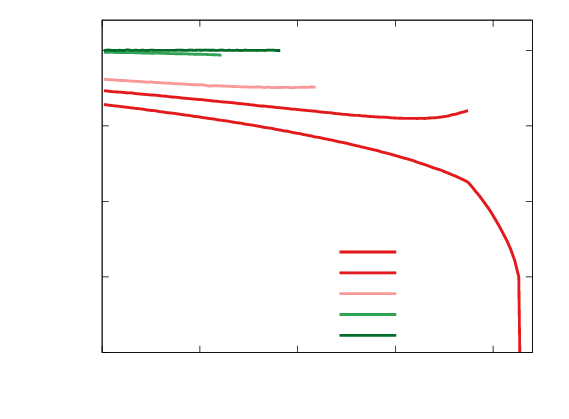}}\gplfronttext
  \end{picture}\endgroup
 }\resizebox{!}{0.35\textwidth}{\begingroup
  \makeatletter
  \providecommand\color[2][]{\GenericError{(gnuplot) \space\space\space\@spaces}{Package color not loaded in conjunction with
      terminal option `colourtext'}{See the gnuplot documentation for explanation.}{Either use 'blacktext' in gnuplot or load the package
      color.sty in LaTeX.}\renewcommand\color[2][]{}}\providecommand\includegraphics[2][]{\GenericError{(gnuplot) \space\space\space\@spaces}{Package graphicx or graphics not loaded}{See the gnuplot documentation for explanation.}{The gnuplot epslatex terminal needs graphicx.sty or graphics.sty.}\renewcommand\includegraphics[2][]{}}\providecommand\rotatebox[2]{#2}\@ifundefined{ifGPcolor}{\newif\ifGPcolor
    \GPcolorfalse
  }{}\@ifundefined{ifGPblacktext}{\newif\ifGPblacktext
    \GPblacktexttrue
  }{}\let\gplgaddtomacro\g@addto@macro
\gdef\gplbacktext{}\gdef\gplfronttext{}\makeatother
  \ifGPblacktext
\def\colorrgb#1{}\def\colorgray#1{}\else
\ifGPcolor
      \def\colorrgb#1{\color[rgb]{#1}}\def\colorgray#1{\color[gray]{#1}}\expandafter\def\csname LTw\endcsname{\color{white}}\expandafter\def\csname LTb\endcsname{\color{black}}\expandafter\def\csname LTa\endcsname{\color{black}}\expandafter\def\csname LT0\endcsname{\color[rgb]{1,0,0}}\expandafter\def\csname LT1\endcsname{\color[rgb]{0,1,0}}\expandafter\def\csname LT2\endcsname{\color[rgb]{0,0,1}}\expandafter\def\csname LT3\endcsname{\color[rgb]{1,0,1}}\expandafter\def\csname LT4\endcsname{\color[rgb]{0,1,1}}\expandafter\def\csname LT5\endcsname{\color[rgb]{1,1,0}}\expandafter\def\csname LT6\endcsname{\color[rgb]{0,0,0}}\expandafter\def\csname LT7\endcsname{\color[rgb]{1,0.3,0}}\expandafter\def\csname LT8\endcsname{\color[rgb]{0.5,0.5,0.5}}\else
\def\colorrgb#1{\color{black}}\def\colorgray#1{\color[gray]{#1}}\expandafter\def\csname LTw\endcsname{\color{white}}\expandafter\def\csname LTb\endcsname{\color{black}}\expandafter\def\csname LTa\endcsname{\color{black}}\expandafter\def\csname LT0\endcsname{\color{black}}\expandafter\def\csname LT1\endcsname{\color{black}}\expandafter\def\csname LT2\endcsname{\color{black}}\expandafter\def\csname LT3\endcsname{\color{black}}\expandafter\def\csname LT4\endcsname{\color{black}}\expandafter\def\csname LT5\endcsname{\color{black}}\expandafter\def\csname LT6\endcsname{\color{black}}\expandafter\def\csname LT7\endcsname{\color{black}}\expandafter\def\csname LT8\endcsname{\color{black}}\fi
  \fi
    \setlength{\unitlength}{0.0500bp}\ifx\gptboxheight\undefined \newlength{\gptboxheight}\newlength{\gptboxwidth}\newsavebox{\gptboxtext}\fi \setlength{\fboxrule}{0.5pt}\setlength{\fboxsep}{1pt}\begin{picture}(5472.00,4030.00)\gplgaddtomacro\gplbacktext{\csname LTb\endcsname \put(860,1131){\makebox(0,0)[r]{\strut{}54.0}}\put(860,1867){\makebox(0,0)[r]{\strut{}57.0}}\put(860,2602){\makebox(0,0)[r]{\strut{}60.0}}\put(860,3338){\makebox(0,0)[r]{\strut{}63.0}}\put(1466,440){\makebox(0,0){\strut{}0.9}}\put(2195,440){\makebox(0,0){\strut{}1.2}}\put(2924,440){\makebox(0,0){\strut{}1.5}}\put(3653,440){\makebox(0,0){\strut{}1.8}}\put(4382,440){\makebox(0,0){\strut{}2.1}}\put(5111,440){\makebox(0,0){\strut{}2.4}}\put(8,3829){\makebox(0,0)[l]{\strut{}b)}}}\gplgaddtomacro\gplfronttext{\csname LTb\endcsname \put(130,2234){\rotatebox{-270}{\makebox(0,0){\strut{}\large $\theta$}}}\put(3045,140){\makebox(0,0){\strut{}$\rho$}}\csname LTb\endcsname \put(4208,1803){\makebox(0,0)[r]{\strut{}Triangular}}\csname LTb\endcsname \put(4208,1603){\makebox(0,0)[r]{\strut{}Cluster 2}}\csname LTb\endcsname \put(4208,1403){\makebox(0,0)[r]{\strut{}Cluster 2a}}\csname LTb\endcsname \put(4208,1203){\makebox(0,0)[r]{\strut{}Cluster 3}}\csname LTb\endcsname \put(4208,1003){\makebox(0,0)[r]{\strut{}Cluster 3a}}\csname LTb\endcsname \put(4208,803){\makebox(0,0)[r]{\strut{}Cluster 4}}}\gplbacktext
    \put(0,0){\includegraphics{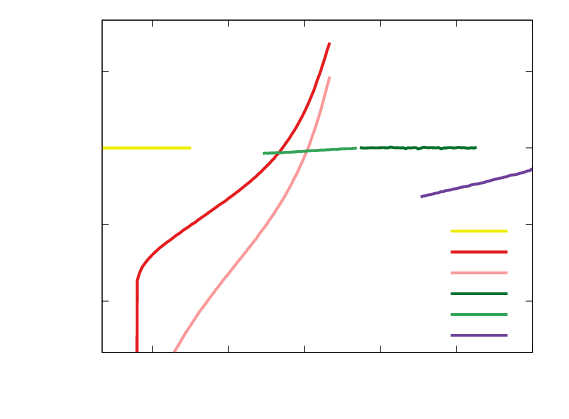}}\gplfronttext
  \end{picture}\endgroup
 }
    \caption{Ground state angle value of the oblique lattice formed by the centroids of clusters. a) as a function of $\ell_C$ for different values of $\rho$, and b) as a function of $\rho$ for $\ell_C=3.99$. The triangular lattice case corresponds to $\theta = 60^\circ$.}
    \label{fig:angle_minimized}
\end{figure}

As discussed previously, cluster anisotropy distorts the lattice, even in the regime when the size of the clusters is relatively small in comparison with the lattice spacing of the cluster lattice. This behavior is however much stronger for more elongated clusters, such as when we have dimer phases. In Fig.~\ref{fig:angle_minimized}, we explore the behavior of the equilibrium $\theta$ angle for different values of $\ell_C$ and $\rho$, characterizing how close to the triangular case ($\theta = 60^\circ$) the general oblique cluster lattice is for the different cluster phases. As can be observed, for the Cluster 3 or 3a phases, where the internal symmetry of the clusters are compatible with the triangular lattice, the lattice angles are very close to $60^\circ$. On the other hand, for the Cluster 2 or 2a phases, the distortion is larger and converges very slowly to $60^\circ$ as $ C\rightarrow 0$ ($\ell_C \to \infty$). This slow convergence indicate, as mentioned before, that even for quite small clusters the equilibrium cluster lattice is significantly anisotropic. {It is worth mentioning that this effect is noticeable in the structure factors of the simulation configurations shown in Fig.~\ref{fig:simu1}, indicating that the trimer phases retains its six-fold rotational symmetry more effectively, whereas in dimer phases this symmetry is typically reduced to two-fold.}

\subsection{Coexistence regions}

The phase diagram presented in Fig.~(\ref{fig:diag}) is not thermodynamically accurate as it does not take into account the possibility of phase coexistence as a mechanism to further minimize the energy of the system in the vicinities of discontinuous phase transitions. This aspect is simple to be corrected as the coexistence region associated to a first order phase transition can be calculated using the standard Maxwell construction. By matching the pressure $P$ and chemical potential $\mu$ of the competing phases at the beginning and end of the coexistence region,  one guarantees that the first-order transition occurs through phase separation process rather than through an abrupt homogeneous configuration change. 

The relevant thermodynamic quantities, pressure and chemical potential, can be calculated as
\begin{equation}
P \equiv - \frac{\partial E}{\partial V} = \rho^2 \frac{\partial \varepsilon}{\partial \rho}, \quad
\mu \equiv \frac{\partial E}{\partial N} = \varepsilon + \rho \frac{\partial \varepsilon}{\partial \rho},
\end{equation}
where $\varepsilon\equiv E/N$ stands for the energy per particle function, which is determined from the ground state energy calculation. {To obtain the pressure and chemical potential, which require the derivative of $\varepsilon(\rho)$, we use a cubic spline interpolation of the discrete energy data. The coexistence boundaries $(\rho_1, \rho_2)$ are then found by numerically solving the following coupled equations using a two-dimensional bisection method~\cite{SciPy2020}}:
\begin{equation}
\left\{
\begin{aligned}
P_A(\rho_1) &= P_B(\rho_2), \\
\mu_A(\rho_1) &= \mu_B(\rho_2),
\end{aligned}
\right.
\end{equation}
where $\rho_1$ and $\rho_2$ marks the boundaries of the coexistence region along the density axis. The energy per particle through the coexistence region can then be evaluated by integrating the equation for the pressure, which yields  
\begin{equation}
\varepsilon_\text{coex}(\rho) = \varepsilon(\rho_1) + P \left( \frac{1}{\rho_1} - \frac{1}{\rho} \right),
\label{eq:enecoex}
\end{equation}
allowing a consistent energetic comparison with the pure phases. 

\begin{figure}[ht!]
    \centering
    \begingroup
  \makeatletter
  \providecommand\color[2][]{\GenericError{(gnuplot) \space\space\space\@spaces}{Package color not loaded in conjunction with
      terminal option `colourtext'}{See the gnuplot documentation for explanation.}{Either use 'blacktext' in gnuplot or load the package
      color.sty in LaTeX.}\renewcommand\color[2][]{}}\providecommand\includegraphics[2][]{\GenericError{(gnuplot) \space\space\space\@spaces}{Package graphicx or graphics not loaded}{See the gnuplot documentation for explanation.}{The gnuplot epslatex terminal needs graphicx.sty or graphics.sty.}\renewcommand\includegraphics[2][]{}}\providecommand\rotatebox[2]{#2}\@ifundefined{ifGPcolor}{\newif\ifGPcolor
    \GPcolorfalse
  }{}\@ifundefined{ifGPblacktext}{\newif\ifGPblacktext
    \GPblacktexttrue
  }{}\let\gplgaddtomacro\g@addto@macro
\gdef\gplbacktext{}\gdef\gplfronttext{}\makeatother
  \ifGPblacktext
\def\colorrgb#1{}\def\colorgray#1{}\else
\ifGPcolor
      \def\colorrgb#1{\color[rgb]{#1}}\def\colorgray#1{\color[gray]{#1}}\expandafter\def\csname LTw\endcsname{\color{white}}\expandafter\def\csname LTb\endcsname{\color{black}}\expandafter\def\csname LTa\endcsname{\color{black}}\expandafter\def\csname LT0\endcsname{\color[rgb]{1,0,0}}\expandafter\def\csname LT1\endcsname{\color[rgb]{0,1,0}}\expandafter\def\csname LT2\endcsname{\color[rgb]{0,0,1}}\expandafter\def\csname LT3\endcsname{\color[rgb]{1,0,1}}\expandafter\def\csname LT4\endcsname{\color[rgb]{0,1,1}}\expandafter\def\csname LT5\endcsname{\color[rgb]{1,1,0}}\expandafter\def\csname LT6\endcsname{\color[rgb]{0,0,0}}\expandafter\def\csname LT7\endcsname{\color[rgb]{1,0.3,0}}\expandafter\def\csname LT8\endcsname{\color[rgb]{0.5,0.5,0.5}}\else
\def\colorrgb#1{\color{black}}\def\colorgray#1{\color[gray]{#1}}\expandafter\def\csname LTw\endcsname{\color{white}}\expandafter\def\csname LTb\endcsname{\color{black}}\expandafter\def\csname LTa\endcsname{\color{black}}\expandafter\def\csname LT0\endcsname{\color{black}}\expandafter\def\csname LT1\endcsname{\color{black}}\expandafter\def\csname LT2\endcsname{\color{black}}\expandafter\def\csname LT3\endcsname{\color{black}}\expandafter\def\csname LT4\endcsname{\color{black}}\expandafter\def\csname LT5\endcsname{\color{black}}\expandafter\def\csname LT6\endcsname{\color{black}}\expandafter\def\csname LT7\endcsname{\color{black}}\expandafter\def\csname LT8\endcsname{\color{black}}\fi
  \fi
    \setlength{\unitlength}{0.0500bp}\ifx\gptboxheight\undefined \newlength{\gptboxheight}\newlength{\gptboxwidth}\newsavebox{\gptboxtext}\fi \setlength{\fboxrule}{0.5pt}\setlength{\fboxsep}{1pt}\begin{picture}(6048.00,4030.00)\gplgaddtomacro\gplbacktext{\csname LTb\endcsname \put(740,3829){\makebox(0,0)[r]{\strut{}0.5}}\put(740,3373){\makebox(0,0)[r]{\strut{}1.0}}\put(740,2918){\makebox(0,0)[r]{\strut{}1.5}}\put(740,2462){\makebox(0,0)[r]{\strut{}2.0}}\put(740,2007){\makebox(0,0)[r]{\strut{}2.5}}\put(740,1551){\makebox(0,0)[r]{\strut{}3.0}}\put(740,1096){\makebox(0,0)[r]{\strut{}3.5}}\put(740,640){\makebox(0,0)[r]{\strut{}4.0}}\put(1266,440){\makebox(0,0){\strut{}0.9}}\put(1875,440){\makebox(0,0){\strut{}1.2}}\put(2484,440){\makebox(0,0){\strut{}1.5}}\put(3093,440){\makebox(0,0){\strut{}1.8}}\put(3702,440){\makebox(0,0){\strut{}2.1}}\put(4311,440){\makebox(0,0){\strut{}2.4}}}\gplgaddtomacro\gplfronttext{\csname LTb\endcsname \put(130,2234){\rotatebox{-270}{\makebox(0,0){\strut{}$\ell_C$}}}\put(2636,140){\makebox(0,0){\strut{}$\rho$}}\csname LTb\endcsname \put(740,3829){\makebox(0,0)[r]{\strut{}0.5}}\put(740,3373){\makebox(0,0)[r]{\strut{}1.0}}\put(740,2918){\makebox(0,0)[r]{\strut{}1.5}}\put(740,2462){\makebox(0,0)[r]{\strut{}2.0}}\put(740,2007){\makebox(0,0)[r]{\strut{}2.5}}\put(740,1551){\makebox(0,0)[r]{\strut{}3.0}}\put(740,1096){\makebox(0,0)[r]{\strut{}3.5}}\put(740,640){\makebox(0,0)[r]{\strut{}4.0}}\put(1266,440){\makebox(0,0){\strut{}0.9}}\put(1875,440){\makebox(0,0){\strut{}1.2}}\put(2484,440){\makebox(0,0){\strut{}1.5}}\put(3093,440){\makebox(0,0){\strut{}1.8}}\put(3702,440){\makebox(0,0){\strut{}2.1}}\put(4311,440){\makebox(0,0){\strut{}2.4}}\put(4798,772){\makebox(0,0)[l]{\strut{}Cluster 4}}\put(4798,1038){\makebox(0,0)[l]{\strut{}Cluster 3a}}\put(4798,1304){\makebox(0,0)[l]{\strut{}Cluster 3}}\put(4798,1570){\makebox(0,0)[l]{\strut{}Cluster 2a}}\put(4798,1835){\makebox(0,0)[l]{\strut{}Cluster 2}}\put(4798,2101){\makebox(0,0)[l]{\strut{}Kagome}}\put(4798,2367){\makebox(0,0)[l]{\strut{}Honeycomb}}\put(4798,2633){\makebox(0,0)[l]{\strut{}Oblique}}\put(4798,2898){\makebox(0,0)[l]{\strut{}Rectangular}}\put(4798,3164){\makebox(0,0)[l]{\strut{}Square}}\put(4798,3430){\makebox(0,0)[l]{\strut{}Triangular}}\put(4798,3696){\makebox(0,0)[l]{\strut{}Coexistences}}}\gplbacktext
    \put(0,0){\includegraphics{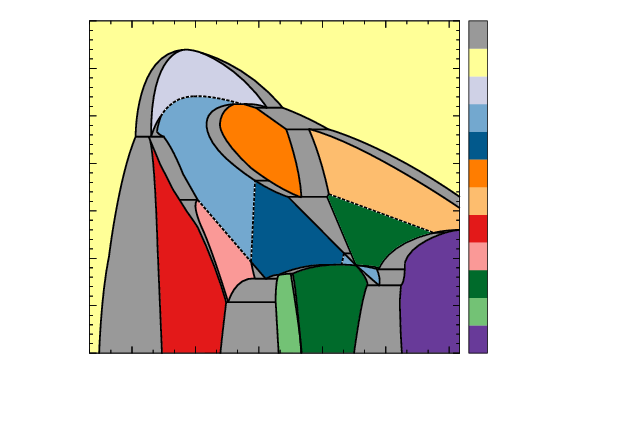}}\gplfronttext
  \end{picture}\endgroup
     \caption{Ground-state phase diagram including coexistence between phases across first-order transitions. Grey regions represent the coexistence between neighboring phases along the density axis.}
    \label{fig:coex_diag}
\end{figure}

In Figure \ref{fig:coex_diag} we show the resulting phase diagram once we take into account the presence of coexistence regions. As can be observed this improvement modifies significantly the thermodynamic phase diagram of the system, suppressing or eliminating entirely many of the pure phases in our system. This is the case, for instance, of the Cluster~2, Cluster~2a and Cluster~3a phases which presents significant coexistence regions at their phase boundaries. Moreover, we can verify the thermodynamic consistency of our calculations by examining the continuity of the extension of the coexistence regions when bifurcations in these regions occurs varying $\ell_C$. It is worth noticing that the extension of a coexistence region is an indicator of how strong a first order transition is, or in other words, how different the curves $\varepsilon(\rho)$ behave for each phase in the vicinity of the crossing density. Interestingly, the lower the values of $C$ the stronger are the first order transitions developed by the model. This indicates that the discontinuous transitions between configurations with different cluster occupancies are stronger than those in which single particle lattices reorganize in a different structure.

\begin{figure}[ht!]
    \centering
    \resizebox{!}{.27\textwidth}{\begingroup
  \makeatletter
  \providecommand\color[2][]{\GenericError{(gnuplot) \space\space\space\@spaces}{Package color not loaded in conjunction with
      terminal option `colourtext'}{See the gnuplot documentation for explanation.}{Either use 'blacktext' in gnuplot or load the package
      color.sty in LaTeX.}\renewcommand\color[2][]{}}\providecommand\includegraphics[2][]{\GenericError{(gnuplot) \space\space\space\@spaces}{Package graphicx or graphics not loaded}{See the gnuplot documentation for explanation.}{The gnuplot epslatex terminal needs graphicx.sty or graphics.sty.}\renewcommand\includegraphics[2][]{}}\providecommand\rotatebox[2]{#2}\@ifundefined{ifGPcolor}{\newif\ifGPcolor
    \GPcolorfalse
  }{}\@ifundefined{ifGPblacktext}{\newif\ifGPblacktext
    \GPblacktexttrue
  }{}\let\gplgaddtomacro\g@addto@macro
\gdef\gplbacktext{}\gdef\gplfronttext{}\makeatother
  \ifGPblacktext
\def\colorrgb#1{}\def\colorgray#1{}\else
\ifGPcolor
      \def\colorrgb#1{\color[rgb]{#1}}\def\colorgray#1{\color[gray]{#1}}\expandafter\def\csname LTw\endcsname{\color{white}}\expandafter\def\csname LTb\endcsname{\color{black}}\expandafter\def\csname LTa\endcsname{\color{black}}\expandafter\def\csname LT0\endcsname{\color[rgb]{1,0,0}}\expandafter\def\csname LT1\endcsname{\color[rgb]{0,1,0}}\expandafter\def\csname LT2\endcsname{\color[rgb]{0,0,1}}\expandafter\def\csname LT3\endcsname{\color[rgb]{1,0,1}}\expandafter\def\csname LT4\endcsname{\color[rgb]{0,1,1}}\expandafter\def\csname LT5\endcsname{\color[rgb]{1,1,0}}\expandafter\def\csname LT6\endcsname{\color[rgb]{0,0,0}}\expandafter\def\csname LT7\endcsname{\color[rgb]{1,0.3,0}}\expandafter\def\csname LT8\endcsname{\color[rgb]{0.5,0.5,0.5}}\else
\def\colorrgb#1{\color{black}}\def\colorgray#1{\color[gray]{#1}}\expandafter\def\csname LTw\endcsname{\color{white}}\expandafter\def\csname LTb\endcsname{\color{black}}\expandafter\def\csname LTa\endcsname{\color{black}}\expandafter\def\csname LT0\endcsname{\color{black}}\expandafter\def\csname LT1\endcsname{\color{black}}\expandafter\def\csname LT2\endcsname{\color{black}}\expandafter\def\csname LT3\endcsname{\color{black}}\expandafter\def\csname LT4\endcsname{\color{black}}\expandafter\def\csname LT5\endcsname{\color{black}}\expandafter\def\csname LT6\endcsname{\color{black}}\expandafter\def\csname LT7\endcsname{\color{black}}\expandafter\def\csname LT8\endcsname{\color{black}}\fi
  \fi
    \setlength{\unitlength}{0.0500bp}\ifx\gptboxheight\undefined \newlength{\gptboxheight}\newlength{\gptboxwidth}\newsavebox{\gptboxtext}\fi \setlength{\fboxrule}{0.5pt}\setlength{\fboxsep}{1pt}\begin{picture}(11664.00,3240.00)\gplgaddtomacro\gplbacktext{\csname LTb\endcsname \put(740,640){\makebox(0,0)[r]{\strut{}0.0}}\put(740,1120){\makebox(0,0)[r]{\strut{}0.3}}\put(740,1600){\makebox(0,0)[r]{\strut{}0.6}}\put(740,2079){\makebox(0,0)[r]{\strut{}0.9}}\put(740,2559){\makebox(0,0)[r]{\strut{}1.2}}\put(740,3039){\makebox(0,0)[r]{\strut{}1.5}}\put(860,440){\makebox(0,0){\strut{}0.70}}\put(1527,440){\makebox(0,0){\strut{}0.90}}\put(2194,440){\makebox(0,0){\strut{}1.10}}\put(2860,440){\makebox(0,0){\strut{}1.30}}\put(3527,440){\makebox(0,0){\strut{}1.50}}}\gplgaddtomacro\gplfronttext{\csname LTb\endcsname \put(190,1839){\rotatebox{-270}{\makebox(0,0){\strut{}$\varepsilon$}}}\put(2193,140){\makebox(0,0){\strut{}$\rho$}}}\gplgaddtomacro\gplbacktext{\csname LTb\endcsname \put(4628,640){\makebox(0,0)[r]{\strut{}0.6}}\put(4628,1120){\makebox(0,0)[r]{\strut{}1.2}}\put(4628,1600){\makebox(0,0)[r]{\strut{}1.8}}\put(4628,2079){\makebox(0,0)[r]{\strut{}2.4}}\put(4628,2559){\makebox(0,0)[r]{\strut{}3.0}}\put(4628,3039){\makebox(0,0)[r]{\strut{}3.6}}\put(4748,440){\makebox(0,0){\strut{}0.70}}\put(5415,440){\makebox(0,0){\strut{}0.90}}\put(6082,440){\makebox(0,0){\strut{}1.10}}\put(6748,440){\makebox(0,0){\strut{}1.30}}\put(7415,440){\makebox(0,0){\strut{}1.50}}}\gplgaddtomacro\gplfronttext{{\small
    \csname LTb\endcsname \put(4078,1839){\rotatebox{-270}{\makebox(0,0){\strut{}$P$}}}\put(6081,140){\makebox(0,0){\strut{}$\rho$}}\csname LTb\endcsname \put(6105,2880){\makebox(0,0)[r]{\strut{}Triangular}}\put(6105,2690){\makebox(0,0)[r]{\strut{}Cluster 2}}\put(6105,2500){\makebox(0,0)[r]{\strut{}Cluster 2a}}\put(6105,2310){\makebox(0,0)[r]{\strut{}Rectangular}}\put(6105,2120){\makebox(0,0)[r]{\strut{}Coexistences}}}}
    \gplgaddtomacro\gplbacktext{\csname LTb\endcsname \put(8516,1087){\makebox(0,0)[r]{\strut{}1.8}}\put(8516,1575){\makebox(0,0)[r]{\strut{}2.4}}\put(8516,2063){\makebox(0,0)[r]{\strut{}3.0}}\put(8516,2551){\makebox(0,0)[r]{\strut{}3.6}}\put(8516,3039){\makebox(0,0)[r]{\strut{}4.2}}\put(8636,440){\makebox(0,0){\strut{}0.70}}\put(9303,440){\makebox(0,0){\strut{}0.90}}\put(9970,440){\makebox(0,0){\strut{}1.10}}\put(10636,440){\makebox(0,0){\strut{}1.30}}\put(11303,440){\makebox(0,0){\strut{}1.50}}}\gplgaddtomacro\gplfronttext{\csname LTb\endcsname \put(7966,1839){\rotatebox{-270}{\makebox(0,0){\strut{}$\mu$}}}\put(9969,140){\makebox(0,0){\strut{}$\rho$}}}\gplgaddtomacro\gplbacktext{\csname LTb\endcsname \put(2234,818){\makebox(0,0)[r]{\strut{}0.96}}\put(2234,1184){\makebox(0,0)[r]{\strut{}1.00}}\put(2234,1550){\makebox(0,0)[r]{\strut{}1.04}}\put(2354,1841){\makebox(0,0){\strut{}1.15}}\put(2834,1841){\makebox(0,0){\strut{}1.20}}\put(3313,1841){\makebox(0,0){\strut{}1.25}}}\gplgaddtomacro\gplfronttext{}\gplbacktext
    \put(0,0){\includegraphics{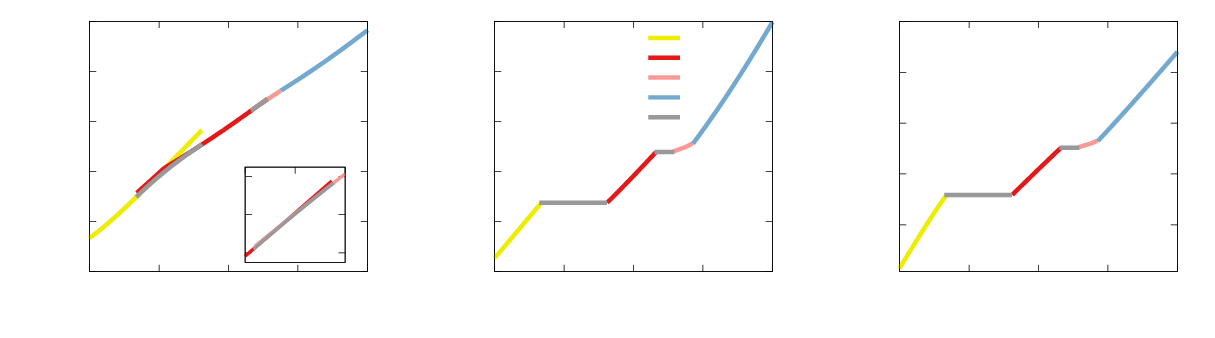}}\gplfronttext
  \end{picture}\endgroup
 }
    \resizebox{!}{.27\textwidth}{\begingroup
  \makeatletter
  \providecommand\color[2][]{\GenericError{(gnuplot) \space\space\space\@spaces}{Package color not loaded in conjunction with
      terminal option `colourtext'}{See the gnuplot documentation for explanation.}{Either use 'blacktext' in gnuplot or load the package
      color.sty in LaTeX.}\renewcommand\color[2][]{}}\providecommand\includegraphics[2][]{\GenericError{(gnuplot) \space\space\space\@spaces}{Package graphicx or graphics not loaded}{See the gnuplot documentation for explanation.}{The gnuplot epslatex terminal needs graphicx.sty or graphics.sty.}\renewcommand\includegraphics[2][]{}}\providecommand\rotatebox[2]{#2}\@ifundefined{ifGPcolor}{\newif\ifGPcolor
    \GPcolorfalse
  }{}\@ifundefined{ifGPblacktext}{\newif\ifGPblacktext
    \GPblacktexttrue
  }{}\let\gplgaddtomacro\g@addto@macro
\gdef\gplbacktext{}\gdef\gplfronttext{}\makeatother
  \ifGPblacktext
\def\colorrgb#1{}\def\colorgray#1{}\else
\ifGPcolor
      \def\colorrgb#1{\color[rgb]{#1}}\def\colorgray#1{\color[gray]{#1}}\expandafter\def\csname LTw\endcsname{\color{white}}\expandafter\def\csname LTb\endcsname{\color{black}}\expandafter\def\csname LTa\endcsname{\color{black}}\expandafter\def\csname LT0\endcsname{\color[rgb]{1,0,0}}\expandafter\def\csname LT1\endcsname{\color[rgb]{0,1,0}}\expandafter\def\csname LT2\endcsname{\color[rgb]{0,0,1}}\expandafter\def\csname LT3\endcsname{\color[rgb]{1,0,1}}\expandafter\def\csname LT4\endcsname{\color[rgb]{0,1,1}}\expandafter\def\csname LT5\endcsname{\color[rgb]{1,1,0}}\expandafter\def\csname LT6\endcsname{\color[rgb]{0,0,0}}\expandafter\def\csname LT7\endcsname{\color[rgb]{1,0.3,0}}\expandafter\def\csname LT8\endcsname{\color[rgb]{0.5,0.5,0.5}}\else
\def\colorrgb#1{\color{black}}\def\colorgray#1{\color[gray]{#1}}\expandafter\def\csname LTw\endcsname{\color{white}}\expandafter\def\csname LTb\endcsname{\color{black}}\expandafter\def\csname LTa\endcsname{\color{black}}\expandafter\def\csname LT0\endcsname{\color{black}}\expandafter\def\csname LT1\endcsname{\color{black}}\expandafter\def\csname LT2\endcsname{\color{black}}\expandafter\def\csname LT3\endcsname{\color{black}}\expandafter\def\csname LT4\endcsname{\color{black}}\expandafter\def\csname LT5\endcsname{\color{black}}\expandafter\def\csname LT6\endcsname{\color{black}}\expandafter\def\csname LT7\endcsname{\color{black}}\expandafter\def\csname LT8\endcsname{\color{black}}\fi
  \fi
    \setlength{\unitlength}{0.0500bp}\ifx\gptboxheight\undefined \newlength{\gptboxheight}\newlength{\gptboxwidth}\newsavebox{\gptboxtext}\fi \setlength{\fboxrule}{0.5pt}\setlength{\fboxsep}{1pt}\begin{picture}(11664.00,3240.00)\gplgaddtomacro\gplbacktext{\csname LTb\endcsname \put(740,640){\makebox(0,0)[r]{\strut{}0.0}}\put(740,1120){\makebox(0,0)[r]{\strut{}0.5}}\put(740,1600){\makebox(0,0)[r]{\strut{}1.0}}\put(740,2079){\makebox(0,0)[r]{\strut{}1.5}}\put(740,2559){\makebox(0,0)[r]{\strut{}2.0}}\put(740,3039){\makebox(0,0)[r]{\strut{}2.5}}\put(860,440){\makebox(0,0){\strut{}0.7}}\put(1488,440){\makebox(0,0){\strut{}1.1}}\put(2115,440){\makebox(0,0){\strut{}1.5}}\put(2743,440){\makebox(0,0){\strut{}1.9}}\put(3370,440){\makebox(0,0){\strut{}2.3}}}\gplgaddtomacro\gplfronttext{\csname LTb\endcsname \put(190,1839){\rotatebox{-270}{\makebox(0,0){\strut{}$\varepsilon$}}}\put(2193,140){\makebox(0,0){\strut{}$\rho$}}}\gplgaddtomacro\gplbacktext{\csname LTb\endcsname \put(4628,640){\makebox(0,0)[r]{\strut{}0.0}}\put(4628,1204){\makebox(0,0)[r]{\strut{}2.0}}\put(4628,1769){\makebox(0,0)[r]{\strut{}4.0}}\put(4628,2333){\makebox(0,0)[r]{\strut{}6.0}}\put(4628,2898){\makebox(0,0)[r]{\strut{}8.0}}\put(4748,440){\makebox(0,0){\strut{}0.7}}\put(5376,440){\makebox(0,0){\strut{}1.1}}\put(6003,440){\makebox(0,0){\strut{}1.5}}\put(6631,440){\makebox(0,0){\strut{}1.9}}\put(7258,440){\makebox(0,0){\strut{}2.3}}}\gplgaddtomacro\gplfronttext{{\small
      \csname LTb\endcsname \put(4078,1839){\rotatebox{-270}{\makebox(0,0){\strut{}$P$}}}\put(6081,140){\makebox(0,0){\strut{}$\rho$}}\csname LTb\endcsname \put(6056,2864){\makebox(0,0)[r]{\strut{}Triangular}}\csname LTb\endcsname \put(6056,2684){\makebox(0,0)[r]{\strut{}Cluster 2}}\csname LTb\endcsname \put(6056,2504){\makebox(0,0)[r]{\strut{}Cluster 3a}}\csname LTb\endcsname \put(6056,2324){\makebox(0,0)[r]{\strut{}Cluster 3}}\csname LTb\endcsname \put(6056,2144){\makebox(0,0)[r]{\strut{}Cluster 4}}\csname LTb\endcsname \put(6056,1964){\makebox(0,0)[r]{\strut{}Coexistences}}}
    }\gplgaddtomacro\gplbacktext{\csname LTb\endcsname \put(8516,640){\makebox(0,0)[r]{\strut{}1.0}}\put(8516,1120){\makebox(0,0)[r]{\strut{}2.0}}\put(8516,1600){\makebox(0,0)[r]{\strut{}3.0}}\put(8516,2079){\makebox(0,0)[r]{\strut{}4.0}}\put(8516,2559){\makebox(0,0)[r]{\strut{}5.0}}\put(8516,3039){\makebox(0,0)[r]{\strut{}6.0}}\put(8636,440){\makebox(0,0){\strut{}0.7}}\put(9264,440){\makebox(0,0){\strut{}1.1}}\put(9891,440){\makebox(0,0){\strut{}1.5}}\put(10519,440){\makebox(0,0){\strut{}1.9}}\put(11146,440){\makebox(0,0){\strut{}2.3}}}\gplgaddtomacro\gplfronttext{\csname LTb\endcsname \put(7966,1839){\rotatebox{-270}{\makebox(0,0){\strut{}$\mu$}}}\put(9969,140){\makebox(0,0){\strut{}$\rho$}}}\gplbacktext
    \put(0,0){\includegraphics{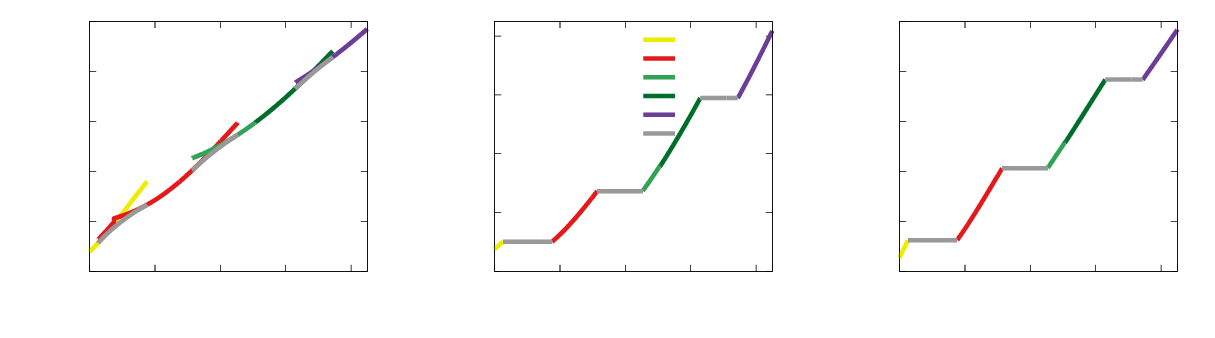}}\gplfronttext
  \end{picture}\endgroup
 }
    \caption{Energy per particle, pressure and chemical potential for different values of $\ell_C$ as a function of $\rho$. First and second row corresponds to $\ell_C=2.51$ and $\ell_C=3.99$, respectively.}
    \label{fig:coex_energy}
\end{figure}

To complement the results for the phase diagram presented we show in Fig.~\ref{fig:coex_energy} the behavior of the energy per particle, pressure and chemical potential increasing density at specific values of $\ell_C$. {The energy per particle as a function of density $\epsilon(\rho)$ is given by the lower envelope of the curves shown on the left column of this figure, where the colored curves correspond to pure phases and the gray curve corresponds to the energy per particle of the coexistence of neighboring phases given by eq.~(\ref{eq:enecoex}). Values of $\epsilon$ higher than the lower envelope indicate metastability of pure phases.} A closer look at this figure reveals the characteristic plateaus in the pressure and chemical potential expected in first order transitions. These plateaus are produced due to the presence of phase-separated domains, where the system locally minimizes its energy while maintaining mechanical and chemical equilibrium. Notably, the widths of most coexistence regions ($\Delta\rho$) are significant, suggesting that the transitions, while discontinuous, do not require extremely fine density control for experimental realization. 

Physically, the coexistence region tends to appear between phases with strongly mismatched local motifs, where continuous parameter deformations are energetically costly or forbidden. For example, coexistence is prominent between Cluster~2 and Cluster~3a, or between Cluster~3 and Cluster~4, whereas transitions involving Oblique and Rectangular structures have narrower coexistence intervals due to the possibility of smooth lattice deformations. {Thermodynamically, continuous transitions correspond to points where the energy-per-particle curves of the two phases meet with the same first derivative with respect to density, as can be seen for instance in Fig.~\ref{fig:coex_energy} in the transition between Cluster-2a and Rectangular phases in the case $\ell_C=2.51$. Consequently, the Maxwell construction applied in such transition would consistently yield a coexistence region of zero width.}\\

\subsection{{Quasicrystalline ground-states}}

\begin{figure}[ht!]
    \centering
    \resizebox{!}{.42\textwidth}{\begingroup
  \makeatletter
  \providecommand\color[2][]{\GenericError{(gnuplot) \space\space\space\@spaces}{Package color not loaded in conjunction with
      terminal option `colourtext'}{See the gnuplot documentation for explanation.}{Either use 'blacktext' in gnuplot or load the package
      color.sty in LaTeX.}\renewcommand\color[2][]{}}\providecommand\includegraphics[2][]{\GenericError{(gnuplot) \space\space\space\@spaces}{Package graphicx or graphics not loaded}{See the gnuplot documentation for explanation.}{The gnuplot epslatex terminal needs graphicx.sty or graphics.sty.}\renewcommand\includegraphics[2][]{}}\providecommand\rotatebox[2]{#2}\@ifundefined{ifGPcolor}{\newif\ifGPcolor
    \GPcolortrue
  }{}\@ifundefined{ifGPblacktext}{\newif\ifGPblacktext
    \GPblacktexttrue
  }{}\let\gplgaddtomacro\g@addto@macro
\gdef\gplbacktext{}\gdef\gplfronttext{}\makeatother
  \ifGPblacktext
\def\colorrgb#1{}\def\colorgray#1{}\else
\ifGPcolor
      \def\colorrgb#1{\color[rgb]{#1}}\def\colorgray#1{\color[gray]{#1}}\expandafter\def\csname LTw\endcsname{\color{white}}\expandafter\def\csname LTb\endcsname{\color{black}}\expandafter\def\csname LTa\endcsname{\color{black}}\expandafter\def\csname LT0\endcsname{\color[rgb]{1,0,0}}\expandafter\def\csname LT1\endcsname{\color[rgb]{0,1,0}}\expandafter\def\csname LT2\endcsname{\color[rgb]{0,0,1}}\expandafter\def\csname LT3\endcsname{\color[rgb]{1,0,1}}\expandafter\def\csname LT4\endcsname{\color[rgb]{0,1,1}}\expandafter\def\csname LT5\endcsname{\color[rgb]{1,1,0}}\expandafter\def\csname LT6\endcsname{\color[rgb]{0,0,0}}\expandafter\def\csname LT7\endcsname{\color[rgb]{1,0.3,0}}\expandafter\def\csname LT8\endcsname{\color[rgb]{0.5,0.5,0.5}}\else
\def\colorrgb#1{\color{black}}\def\colorgray#1{\color[gray]{#1}}\expandafter\def\csname LTw\endcsname{\color{white}}\expandafter\def\csname LTb\endcsname{\color{black}}\expandafter\def\csname LTa\endcsname{\color{black}}\expandafter\def\csname LT0\endcsname{\color{black}}\expandafter\def\csname LT1\endcsname{\color{black}}\expandafter\def\csname LT2\endcsname{\color{black}}\expandafter\def\csname LT3\endcsname{\color{black}}\expandafter\def\csname LT4\endcsname{\color{black}}\expandafter\def\csname LT5\endcsname{\color{black}}\expandafter\def\csname LT6\endcsname{\color{black}}\expandafter\def\csname LT7\endcsname{\color{black}}\expandafter\def\csname LT8\endcsname{\color{black}}\fi
  \fi
    \setlength{\unitlength}{0.0500bp}\ifx\gptboxheight\undefined \newlength{\gptboxheight}\newlength{\gptboxwidth}\newsavebox{\gptboxtext}\fi \setlength{\fboxrule}{0.5pt}\setlength{\fboxsep}{1pt}\definecolor{tbcol}{rgb}{1,1,1}\begin{picture}(4320.00,4320.00)\gplgaddtomacro\gplbacktext{\csname LTb\endcsname \put(150,4019){\makebox(0,0)[l]{\strut{}a)}}}\gplgaddtomacro\gplfronttext{\csname LTb\endcsname \put(2129,4019){\makebox(0,0){\strut{}$\rho=1.75$\quad$\ell_C=1.60$\quad$T=0.0016$}}}\gplgaddtomacro\gplfronttext{}\gplbacktext
    \put(0,0){\includegraphics[width={216.00bp},height={216.00bp}]{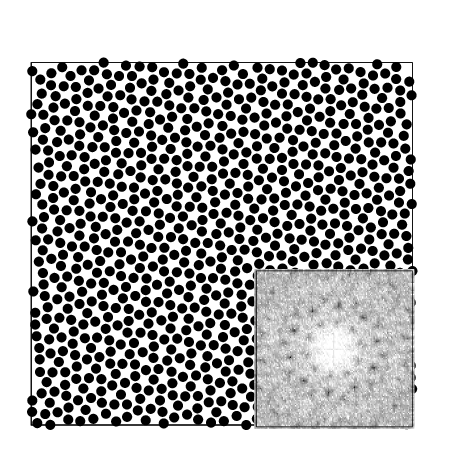}}\gplfronttext
  \end{picture}\endgroup
 }
    \resizebox{!}{.42\textwidth}{\begingroup
  \makeatletter
  \providecommand\color[2][]{\GenericError{(gnuplot) \space\space\space\@spaces}{Package color not loaded in conjunction with
      terminal option `colourtext'}{See the gnuplot documentation for explanation.}{Either use 'blacktext' in gnuplot or load the package
      color.sty in LaTeX.}\renewcommand\color[2][]{}}\providecommand\includegraphics[2][]{\GenericError{(gnuplot) \space\space\space\@spaces}{Package graphicx or graphics not loaded}{See the gnuplot documentation for explanation.}{The gnuplot epslatex terminal needs graphicx.sty or graphics.sty.}\renewcommand\includegraphics[2][]{}}\providecommand\rotatebox[2]{#2}\@ifundefined{ifGPcolor}{\newif\ifGPcolor
    \GPcolortrue
  }{}\@ifundefined{ifGPblacktext}{\newif\ifGPblacktext
    \GPblacktexttrue
  }{}\let\gplgaddtomacro\g@addto@macro
\gdef\gplbacktext{}\gdef\gplfronttext{}\makeatother
  \ifGPblacktext
\def\colorrgb#1{}\def\colorgray#1{}\else
\ifGPcolor
      \def\colorrgb#1{\color[rgb]{#1}}\def\colorgray#1{\color[gray]{#1}}\expandafter\def\csname LTw\endcsname{\color{white}}\expandafter\def\csname LTb\endcsname{\color{black}}\expandafter\def\csname LTa\endcsname{\color{black}}\expandafter\def\csname LT0\endcsname{\color[rgb]{1,0,0}}\expandafter\def\csname LT1\endcsname{\color[rgb]{0,1,0}}\expandafter\def\csname LT2\endcsname{\color[rgb]{0,0,1}}\expandafter\def\csname LT3\endcsname{\color[rgb]{1,0,1}}\expandafter\def\csname LT4\endcsname{\color[rgb]{0,1,1}}\expandafter\def\csname LT5\endcsname{\color[rgb]{1,1,0}}\expandafter\def\csname LT6\endcsname{\color[rgb]{0,0,0}}\expandafter\def\csname LT7\endcsname{\color[rgb]{1,0.3,0}}\expandafter\def\csname LT8\endcsname{\color[rgb]{0.5,0.5,0.5}}\else
\def\colorrgb#1{\color{black}}\def\colorgray#1{\color[gray]{#1}}\expandafter\def\csname LTw\endcsname{\color{white}}\expandafter\def\csname LTb\endcsname{\color{black}}\expandafter\def\csname LTa\endcsname{\color{black}}\expandafter\def\csname LT0\endcsname{\color{black}}\expandafter\def\csname LT1\endcsname{\color{black}}\expandafter\def\csname LT2\endcsname{\color{black}}\expandafter\def\csname LT3\endcsname{\color{black}}\expandafter\def\csname LT4\endcsname{\color{black}}\expandafter\def\csname LT5\endcsname{\color{black}}\expandafter\def\csname LT6\endcsname{\color{black}}\expandafter\def\csname LT7\endcsname{\color{black}}\expandafter\def\csname LT8\endcsname{\color{black}}\fi
  \fi
    \setlength{\unitlength}{0.0500bp}\ifx\gptboxheight\undefined \newlength{\gptboxheight}\newlength{\gptboxwidth}\newsavebox{\gptboxtext}\fi \setlength{\fboxrule}{0.5pt}\setlength{\fboxsep}{1pt}\definecolor{tbcol}{rgb}{1,1,1}\begin{picture}(4320.00,4320.00)\gplgaddtomacro\gplbacktext{\csname LTb\endcsname \put(150,4019){\makebox(0,0)[l]{\strut{}b)}}}\gplgaddtomacro\gplfronttext{\csname LTb\endcsname \put(2129,4019){\makebox(0,0){\strut{}$\rho=1.30$\quad$\ell_C=1.40$\quad$T=0.0015$}}}\gplgaddtomacro\gplfronttext{}\gplbacktext
    \put(0,0){\includegraphics[width={216.00bp},height={216.00bp}]{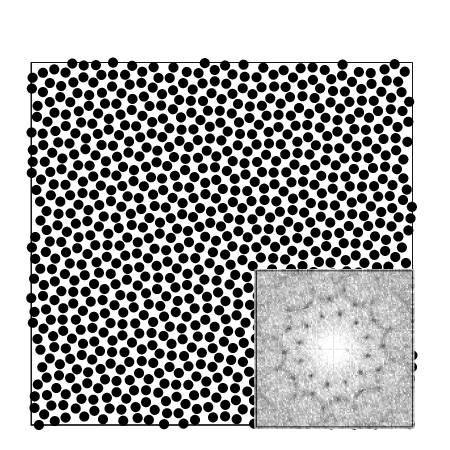}}\gplfronttext
  \end{picture}\endgroup
 }
    \caption{{Decagonal (a) and dodecagonal (b) low temperature molecular dynamics configurations for systems with $N=900$ particles, together with the corresponding values of the density, relative strength of the hard-core interaction and temperature. The associated structure factors of the configurations are shown in the insets confirming their 10-fold or 12-fold rotational symmetry.}}
    \label{fig:simu2}
\end{figure}

\begin{figure}[ht!]
    \centering
\includegraphics[width=.7\textwidth]{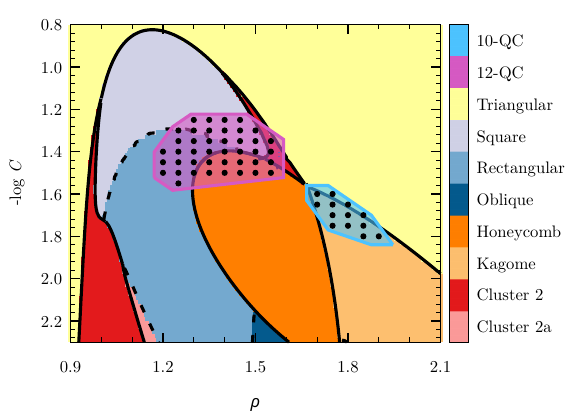}
    \caption{{Phase diagram corrected by molecular dynamics simulations. The highlighted areas are a guide to the eye for the regions where our simulation grid identified the decagonal (10-QC) and dodecagonal (12-QC) quasicrystalline phases. Dots indicate the actual positions where simulations converged to quasicrystalline configurations.}}
    \label{fig:diagsimu}
\end{figure}

{In order to present a more complete picture of the phase diagram presented in Fig.~\ref{fig:diag}, in this section we address the cases where our variational analysis of the ground-states could not capture the behavior observed in our simulations. 
As reported in previous studies \cite{Dotera2014, Zu2017, Dijkstra2017}, two-dimensional systems with HCSS potentials frequently presents quasicrystalline phases in their phase diagrams. In our case, we observe that the simulations discussed in section \ref{model} produce low temperature configurations with decagonal and dodecagonal rotational symmetries, as shown in Fig.~\ref{fig:simu2}. Such configurations are observed in regions of the phase diagram that exhibit a high degree of frustration. The decagonal quasicrystalline configurations are observed in a region near the phase boundaries of the kagome phase with both triangular and honeycomb phases, whereas the dodecagonal quasicrystalline configurations are observed in a lower density region, near the phase boundaries between honeycomb, rectangular and square lattice phases. These regions are highlighted in the ground-state phase diagram shown in Fig.~\ref{fig:diagsimu}.}

{To further validate our characterization of the low-temperature phases, we compared the potential energies obtained from Langevin simulations with those resulting from the minimization of the variational ansatz. For all periodic phases captured by the ansatz, the agreement is excellent, with energy differences well below the numerical resolution of the simulations. Interestingly, the quasicrystalline phases observed in the frustrated region exhibit energies that are extremely close to those of the corresponding minimized ansatz --- typically differing by $\mathcal{O}(10^{-3})$. The quasicrystals display slightly lower energies, suggesting that aperiodic order may outperform the periodic candidates proposed in the variational analysis. This reinforces the view that competing length scales can stabilize nearly degenerate structures.}

\section{Conclusions}
\label{conclusions}

In this work, we have explored the ground-state phase diagram of a two-dimensional core-softened particle system, defined by the combination of the GEM-4 ultra-soft potential and an inverse power-law interaction with a decay exponent of 6 acting as a hard-core repulsion. At first, we conducted an exploration of the phase diagram using both molecular dynamics simulations and numerical energy minimization based on a general ansatz for $n$-occupied cluster phases with an independent orientational order in up to two sublattices. This strategy enabled us to identify the short-wavelength mesophases typically present in this model and to formulate a set of specific ansatz for each phase, respecting their symmetries while relying on the minimal set of parameters and constraints needed for their proper description. 
{While our ansatz set was guided by two exploratory techniques and previously reported phases in similar models, we are aware that more complex periodic phases were observed in similar systems \cite{Zu2017, Dijkstra2017}.
Nevertheless, we are confident that, in this sense, our results captures the relevant physics of this system.}

By minimizing the energy of each phase with respect to its degrees of freedom, using a nonlinear constrained optimization technique, we have constructed the zero-temperature phase diagram using as running parameters the particle density and the relative strength of the hard-core potential compared to the ultra-soft repulsion. The resulting diagram reveals a wide variety of phases emerging in the intermediate regime where both interactions operate on comparable length scales. When the hard-core repulsion is much weaker, we observe cluster crystal phases with increasing occupancy as density grows. We identified that these cluster phases generally adopt oblique lattice arrangements, which tend to evolve toward triangular symmetry as the hard-core repulsion becomes negligible — in other words, as the cluster shape anisotropy becomes less significant due to its smaller size — although this convergence can be slow. A notable exception in this regard is the Cluster 3 phase, where the geometry of each cluster allows it to perfectly match the triangular lattice.

We have identified two distinct $n=2$ cluster (dimer) phases: the Cluster 2 phase, in which dimers align along a line equidistant from neighboring dimers, and the Cluster 2-aligned phase, where dimers are oriented along a primitive direction of the oblique lattice. We did not observe dimer phases with more complex orientational patterns, such as the alternated or “anti-nematic” dimer phases reported previously in similar two-dimensional core-softened systems with triangular-lattice constraints~\cite{Rossini2018, Mambretti2021}. However, we did observe such alternated behavior in the trimer case, the Cluster 3-alternated phase, where the compatibility between the internal cluster symmetry and the triangular lattice permits subtle energy optimization through variations in cluster orientation with minimal lattice distortion.

Beyond the cluster phases, as the hard-core repulsion becomes stronger, we have also identified single-particle phases, including triangular and square lattice arrangements, as well as phases associated with stripe-like mesophases (rectangular and oblique lattices) and hole-like mesophases (kagome and honeycomb lattices). The construction of the specific ansatz and the minimization of their energies not only allowed us to characterize and localize these phases in the phase diagram, but also to calculate the coexistence regions corresponding to the first order phase transitions. 
{Moreover, the fact that we are always locating local energy minima, rather than merely stationary points in the periodic configuration space, already provides strong evidence of the mechanical stability of the obtained phases. Furthermore, since all reported phases were also observed in simulations following careful equilibration and annealing, we can confirm their mechanical stability from a fully general perspective.}

{The rich phase diagram is confirmed through molecular dynamics simulation using a combination of full inertial Langevin dynamics, parallel tempering, and low-temperature annealing. These simulations were essential to prove the existence of quasicrystalline phases in the region of stronger hard-core repulsion of both hole-like mesophases. However, a full integration of these simulation results for quasicrystalline phases with the rest of the phase diagram is beyond the scope of the present study and will be undertaken in a future work.}

{The variety of mesophases found in this work highlights a general principle for systems with competing repulsive interactions at different length scales. While the specific locations of the phase boundaries depend on the details of the pair potential, the overall morphology of the phase diagram and the typical sequence of cluster crystals, stripes, and hole-like phases beneath the reentrant triangular crystal phase are expected to be qualitatively universal.} 

The detailed description of the short-wavelength region of the ground-state phase diagram presented can establish the basis for future investigations of thermal and quantum fluctuation effects on mesophases. For example, this system could be an excellent candidate for studying the melting of anisotropic two-dimensional crystalline phases~\cite{Ostlund1981, Toner1981}, {including the possibility of intermediate phases between the ground-state and liquid phases~\cite{Mendoza-Coto2024}}, or, in the context of interacting bosons, for characterizing the corresponding quantum ground states~\cite{Diaz2015, Lima2025, Cinti2014, Cinti2017, Cinti2019}. Moreover, the classical behavior described here could be extended to higher densities, where longer-wavelength mesophases~\cite{Glaser2007} may give rise to even more complex phenomena than those reported in the present study. {Finally, we hope this study will stimulate further experimental investigations in colloidal systems under 2D confinement \cite{Menath2021, Ciarella2021}.}

\section*{Acknowledgements}

{We thank the computational resources provided by the Laboratório Multiusuário de Pesquisas Físicas (LAMPEF/FSC/CFM/UFSC) and by the Centro Nacional de Processamento de Alto Desempenho em São Paulo (CENAPAD-SP). }

\section*{References}

\providecommand{\newblock}{}

\end{document}